\documentclass[twocolumn,trackchanges,twocolappendix]{aastex7}

\usepackage{natbib}
\usepackage{rotating}
\usepackage{afterpage}
\usepackage{float}

\begin{document}

\title{Redshift Evolution of the Intrinsic Alignments of Early-Type Galaxies and Subhalos in the Horizon Run 5 Simulation}

\correspondingauthor{Motonari Tonegawa}
\email{motonari.tonegawa@apctp.org}

\author[0009-0008-0644-9630]{Sanghyeon Han}
\affiliation{Department of Physics and Astronomy, Seoul National University, Gwanak-gu, Seoul 08826, Republic of Korea}
\email{chzg1@snu.ac.kr}

\author[0000-0001-6400-1692]{Motonari Tonegawa}
\affiliation{Asia Pacific Center for Theoretical Physics, Pohang 37673, Republic of Korea}
\email{motonari.tonegawa@apctp.org}

\author[0000-0003-3428-7612]{Ho Seong Hwang}
\affiliation{Department of Physics and Astronomy, Seoul National University, Gwanak-gu, Seoul 08826, Republic of Korea}
\affiliation{SNU Astronomy Research Center, Seoul National University, Seoul 08826, Republic of Korea}
\affiliation{Australian Astronomical Optics - Macquarie University, 105 Delhi Road, North Ryde, NSW 2113, Australia}
\email{galaxy79@snu.ac.kr}

\author[0000-0003-0225-6387]{Yohan Dubois}
\affiliation{Institut d'Astrophysique de Paris, Sorbonne Université, CNRS, UMR 7095, 98 bis bd Arago, 75014 Paris, France}
\email{dubois@iap.fr}

\author[0000-0002-4391-2275]{Juhan Kim}
\affiliation{Center for Advanced Computation, Korea Institute for Advanced Study, 85 Hoegiro, Dongdaemun-gu, Seoul 02455, Republic of Korea}
\email{kjhan@kias.re.kr}

\author[0000-0003-4164-5414]{Yonghwi Kim}
\affiliation{Korea Institute of Science and Technology Information (KISTI), 245 Daehak-ro, Yuseong-gu, Daejeon 34141, Republic of Korea}
\email{yonghwi.kim@kisti.re.kr}

\author[0000-0001-9734-9257]{Oh-Kyoung Kwon}
\affiliation{Korea Institute of Science and Technology Information (KISTI), 245 Daehak-ro, Yuseong-gu, Daejeon 34141, Republic of Korea}
\email{okkwon@kisti.re.kr}

\author[0000-0002-6810-1778]{Jaehyun Lee}
\affiliation{Korea Astronomy and Space Science Institute, 776, Daedeokdae-ro, Yuseong-gu, Daejeon 34055, Republic of Korea}
\email{jaehyun@kasi.re.kr}

\author{Owain N. Snaith}
\affiliation{Department of Physics and Astronomy, University of Exeter, Exeter EX4 4QL, UK}
\email{o.n.snaith@exeter.ac.uk}


\author[0000-0003-4446-3130]{Brad K. Gibson}
\affiliation{Woodmansey Primary School, Hull Road, Woodmansey HU17 0TH, UK}
\email{brad.k.gibson@gmail.com}

\author[0000-0001-9521-6397]{Changbom Park}
\affiliation{School of Physics, Korea Institute for Advanced Study, 85 Hoegiro, Dongdaemun-gu, Seoul 02455, Republic of Korea}
\email{cbp@kias.re.kr}

\begin{abstract}

We investigate the redshift evolution of intrinsic alignments of the shapes of galaxies and subhalos with the large-scale structures of the universe using the cosmological hydrodynamic simulation, \textit{Horizon Run 5}. To this end, early-type galaxies are selected from the simulated galaxy catalogs based on stellar mass and kinematic morphology. The shapes of galaxies and subhalos are computed using the reduced inertia tensor derived from mass-weighted particle positions. We find that the misalignment between galaxies and their corresponding dark-matter subhalos decreases over time. We further analyze the two-point correlation between galaxy or subhalo shapes and the large-scale density field traced by their spatial distribution, and quantify the amplitude using the nonlinear alignment model across a wide redshift range from $z = 0.625$ to $z = 2.5$. We find that the intrinsic alignment amplitude, $A_{\rm NLA}$, of galaxies remains largely constant with redshift, whereas that of dark matter subhalos exhibits moderate redshift evolution, with a power-law slope that deviates from zero at a significance level exceeding $3\sigma$. Additionally, $A_{\rm NLA}$ is found to depend on both the stellar mass and kinematic morphology of galaxies. 
Notably, our results are broadly consistent with existing observational constraints.
Our findings are in good agreement with previous results of other cosmological simulations.

\end{abstract}



\section{INTRODUCTION} \label{sec:intro}

The shape and spin of galaxies are not randomly oriented but exhibit preferential alignments with the surrounding large-scale structure in the universe.
This intrinsic alignment (IA) of galaxies has come to be considered closely linked to galaxy formation processes and the underlying matter distribution in the universe \citep[for reviews, see][]{Joachimi2015,Kiessling2015,Kirk2015,Troxel2015}.
An early theoretical investigation
was presented in \cite{Pichon1999}; they showed that the kinematics of the large scale flow would impact IAs on scales of  a few $h^{-1}\ \mathrm{Mpc}$ and below. 

On the other hand, the IA of galaxy shapes is a major source of systematic bias in weak gravitational lensing measurements as it produces an effect that closely resembles cosmic shear \citep{Croft2000,Catelan2001,Aubert2004,Hirata2004}. Hence, it significantly contaminates the measure of cosmological parameters, complicating the interpretations of weak lensing results.
Upcoming surveys, such as Euclid \citep{Euclid2024}, the Vera C. Rubin Observatory Legacy Survey of Space and Time \citep{Ivezi2019}, and the Nancy Grace Roman Space Telescope High Latitude Imaging Survey \citep{Akeson2019}, aim to achieve high precision in cosmological measurements from weak lensing observations. 
High-redshift observations, such as those from JWST \citep{Gardner2023}, further highlight the importance of effectively removing the IA contribution for the weak-lensing analysis with the source galaxies over a wide redshift range.
Therefore, understanding and isolating the effect of IA has become an urgent and crucial challenge.

Although intrinsic alignments (IAs) introduce systematic errors in weak-lensing measurements, they have recently been recognized as a valuable tool for cosmological investigations. For instance, IAs can offer insights into primordial non-Gaussianity and enable the extraction of cosmological information from observations of baryon acoustic oscillations \citep[see, e.g.,][]{Chisari2013,Schmidt2015,Okumura2019,Akitsu2021,Kurita2023}. 
\citet{Okumura2022} demonstrated that incorporating intrinsic alignments (IA) and the kinematic Sunyaev-Zel'dovich effect can place more stringent constraints on dark energy and gravity models.

The IA for early-type galaxies is successfully described by the linear alignment (LA) model, which assumes that the intrinsic shape of a galaxy is determined at the epoch of galaxy formation \citep{Hirata2004}. 
In the LA model, the shapes (orientation and ellipticity) of early-type galaxies are aligned in response to the local tidal field through tidal stretching.
In contrast, late-type galaxies acquire the angular momentum under the influence of the tidal field, as described by the tidal torquing mechanism \citep[e.g.,][]{Catelan2001,Crittenden2001,Bailin2005a,Lee2008,Codis2015b}. 
Assuming that their observed ellipticity is determined by the spin orientation, the resulting alignment is modeled as quadratic rather than linear. In this case, the IA signal is expected to be zero in the shape-density correlation, as it is proportional to the cubic of the linear overdensity. 
Indeed, previous studies have confirmed that late-type exhibit null or negligible signals of IAs \citep[e.g.,][]{Mandelbaum2011,Tonegawa2022,Samuroff2023}.

In the LA model, the strength of  IA is quantified by the amplitude parameter, $A_{\rm LA}$, which relates the galaxy shapes to the tidal field induced by the matter density. Observations and simulations suggest that the $A_{\rm{LA}}$ depends on galaxy properties such as color, luminosity, halo mass, and redshift \citep[e.g.,][]{Singh2015,Chisari2016,Piras2018,Johnston2019,Yao2020,Samuroff2021,Fortuna2021,Tonegawa2022,Samuroff2023,Tonegawa2024}. This finding suggests that intrinsic alignments (IA) may result from the formation and evolution of galaxies through their interactions with the large-scale structure. Therefore, a detailed understanding of IA is essential for a comprehensive picture of galaxy formation in the universe.
Moreover, it is important to characterize the redshift evolution of the IAs because recent observations tend to cover a wide range of redshifts, which requires a model for redshift-dependent bias to weak lensing. To summarize, the understanding of the strength and evolution of IA is key to advancing our knowledge of not only the weak lensing but also galaxy formation and evolution.

Previous observations found a nearly constant amplitude of IA up to $z\sim0.7$ \citep{Joachimi2011,Singh2015,Samuroff2023,Hervas-Peters2024} for early-type galaxies.
High-$z$ observations, however, report stronger IA signals than lower redshifts, implying a potential redshift dependence of the IA according to the LA model predictions \citep{Yao2020,Tonegawa2022}. 
Using a dark-matter-only simulation, \citet{Kurita2021} found a redshift evolution in $A_{\rm LA}$ for subhalos. Examining whether a similar trend appears in galaxies of hydrodynamical simulations would be interesting to do. Although less decisive due to the influence of various subgrid physics and numerical solvers, studies based on hydrodynamical simulations seem to show a weaker redshift dependence of the IA \citep{Tenneti2015a,Chisari2016,Samuroff2021}.
The misalignment between galaxies and subhalos may be related to this difference \citep{Okumura2009}.

The goal of this work is to study the redshift dependence of the intrinsic alignment of early-type galaxies and dark-matter subhalos (dark-matter components of galaxies) in the same simulation framework.
We use a cosmological hydrodynamical simulation to measure the IA of early-type galaxies and their host dark-matter subhalos, and investigate their redshift evolution.
We analyze the evolutionary features of IA with respect to redshift, comparing our results with findings of previous works. By examining the redshift dependence, we aim to understand how the evolution and physical properties of galaxies affect their intrinsic alignments. 
As we are interested in the redshift evolution of IAs, we focus on mainly on early-type galaxies, which exhibit measurable and redshift-dependent IA signals.

This paper is structured as follows: In Section 2, we introduce the high-resolution hydrodynamical simulation Horizon Run 5 (HR5) and describe the method selecting early-type galaxies within HR5. Section 3 details the approach for measuring the shapes of galaxies and dark-matter subhalos. We define the shape-density cross-correlation, a two-point statistical method used to measure the IA. This section also outlines the theoretical model of the IA and explains the procedure for estimating its amplitude. In Section 4, we investigate the properties of the sample, and we show the results of the IA measurement and the amplitude for early-type galaxies and dark-matter subhalos. Section 5 focuses on the analysis and discussion of the findings from Section 4, where we compare our results with previous studies and discuss some caveats on the origin of redshift evolution and the time scale of the IA. Finally, we summarize the key findings and propose directions for future research in Section 6. In Appendix, we verify the technical issues on our results.
\begin{figure}[!t]
\centering
\includegraphics[width=0.47\textwidth]{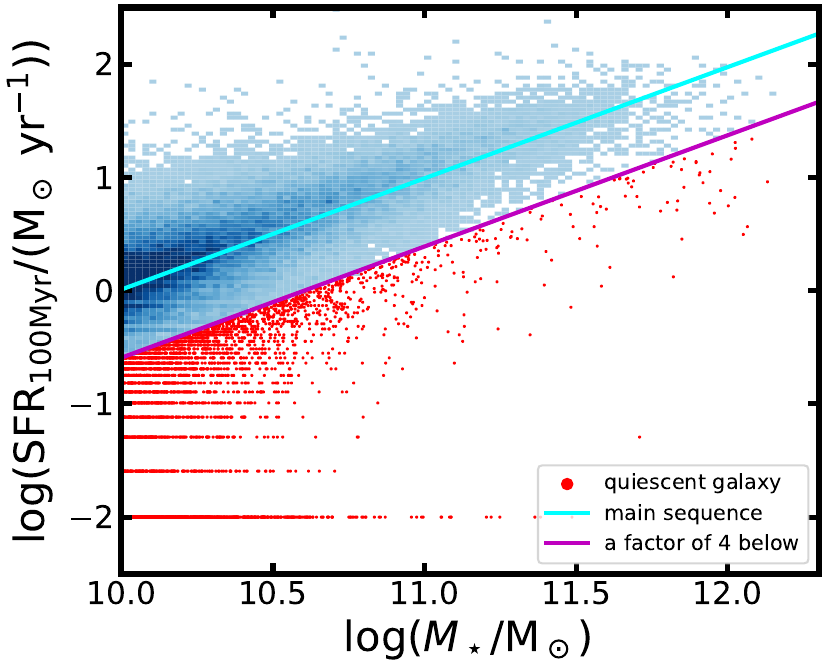}
\caption{A distribution of SFR as a function of stellar mass of HR5 galaxy at $z=0.625$. The value of SFR is averaged over 100 Myr. The bluish boxes are star-forming galaxies. The cyan line is the main sequence of star-forming galaxies and the magenta line marks the level that is one fourth of the main sequence. The red dots are for quiescent galaxies. Note that galaxies with zero SFR at $\log{(\rm{SFR}/(M_{\odot}\ \rm{yr^{-1}}))}=-2$ for the practical display on the logarithmic scale.}
\label{fig:Mstar_SFR}
\end{figure}
\begin{figure}[!t]
\centering
\includegraphics[width=0.47\textwidth]{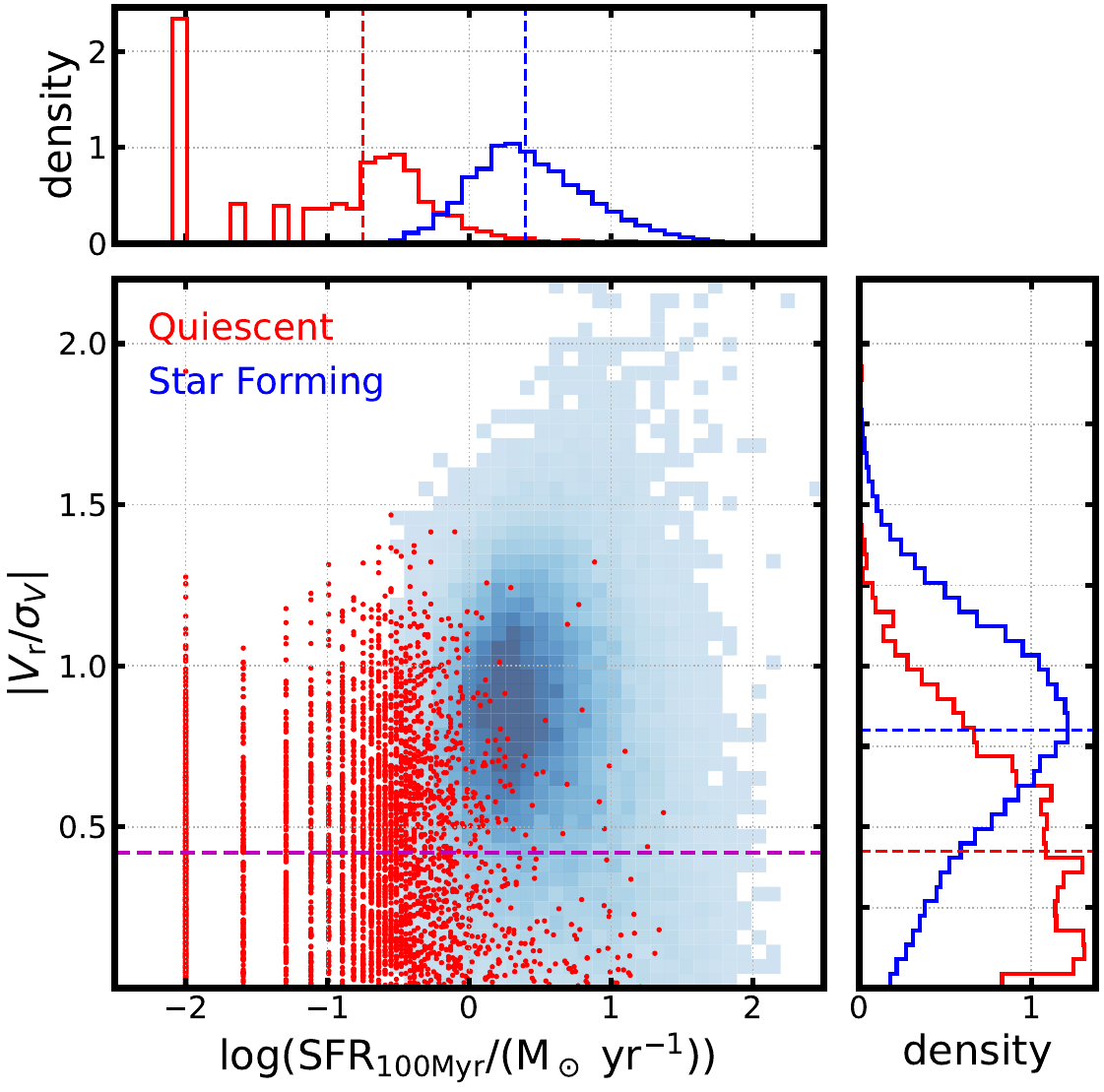}
\caption{A distribution of star forming galaxies and quiescent galaxies on the plane of SFR and $|V_{\rm r}/\sigma_V|$. 
Similar to Figure~\ref{fig:Mstar_SFR}, galaxies with zero SFR are located at $\log{(\rm{SFR}/(M_{\odot}\ \rm{yr^{-1}}))}=-2$. Top and right panels show the normalized histogram of each galaxy type with the median denoted as dashed lines. The magenta dashed line represents the criterion for early-type galaxy  ($|V_{\rm r}/\sigma_V|= 0.42$).}
\label{fig:SFR_ratio}
\end{figure}

\section{DATA} \label{sec:data}

\subsection{Horizon Run 5 Simulation} \label{subsec:HR5}

In this study, we use the high-resolution cosmological hydrodynamical simulation, Horizon Run 5 \citep[HR5;][]{Lee2021}, to investigate the IAs of galaxies and their hosting dark-matter subhalos. HR5 is performed with a modified version of the RAMSES code \citep{Teyssier2002}, which employs an adaptive mesh refinement (AMR) algorithm for higher resolutions in the zoom-in region. The simulation traces the gravitational and hydrodynamic evolution of cosmic matter down to the redshift of $z=0.625$ in a cubic box of a side length, $717\ h^{-1}\ \mathrm{cMpc}$. For a higher-resolution evolution, a volume of a nearly square-rod shape is set as a ``zoom-in" region with dimensions $(717,\ 81,\ 87)\ h^{-1}\ \mathrm{cMpc}$. HR5 maintains a nearly constant resolution down to 1 kpc for the dark matter  and gas.

The simulation adopts cosmological parameters consistent with the Planck 2016 results \citep{Planck2016}, with $\Omega_{\rm{m}}=0.3$, $\Omega_{\Lambda}=0.7$, and $H_0 = 68.4\ \rm{km\ s^{-1}\ Mpc^{-1}}$. Initial conditions are generated using the MUSIC package \citep{Hahn2011}. dark-matter halos are identified using the friends-of-friends algorithm \citep{Davis85,Audit98}, and galaxies are identified with an improved PSB \citep{Kim06}-based galaxy finder, pGalF \citep{Lee2021,Kim2023}. 

All dark-matter particles are identical in mass, whereas the stellar particles exhibit slight variations in mass due to mass loss (by stellar winds and SN explosions) or a different amount gas used for the star formation. In the zoomed region, the mass of the dark-matter particle is $6.89 \times 10^7\ {\rm M}_{\odot}$, while the mass of the stellar particle is on average about $10^6\ {\rm M}_{\odot}$. 

Galaxy merger trees are constructed based on the stellar particles using the tree building algorithm, ySAMtm \citep{Lee2014,Park2022,Kim2023}. For halos with no stellar particles, we trace their progenitors using their most bound dark-matter particles \citep{Hong2016}.
HR5 initially has a based level of 13, which corresponds to 8192 grid cells on a size in the zoomed region, and the grid level decreases to 8 (256 grid cells on a side) in the background volume. 
DM particle mass increases a factor of 8 with each grid level decrease.
Further details on the HR5 simulation can be found in \cite{Lee2021}.
\begin{figure}[!t]
\centering
\includegraphics[width=0.47\textwidth]{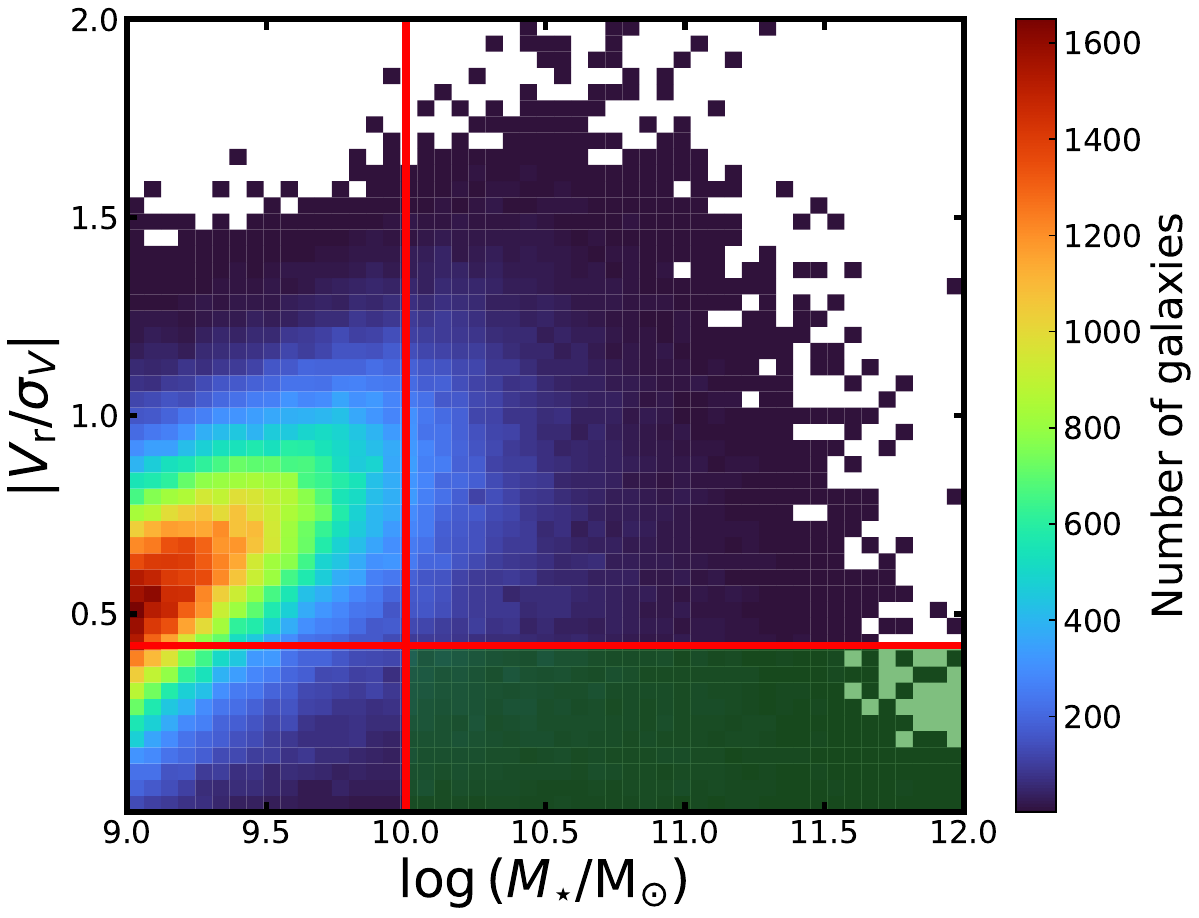}
\caption{A selection plot of early-type galaxies with the stellar mass ($x$--axis) and the kinematic ratio ($y$--axis) at $z=0.625$. The red vertical line indicates the stellar mass cut while the red horizontal line marks the galaxy kinematics cut. The green-shaded region represents the distribution of galaxies that satisfy both criteria and are thus selected for analysis. We use a threshold value of $|V_{\rm r}/\sigma_V|=0.42$, which is determined by the distribution of quiescent galaxies as shown in Figure~\ref{fig:SFR_ratio}. 
\label{fig:target_selection}}
\end{figure}

\subsection{Early-type Galaxy Selection} \label{subsec:target}

We select these galaxies from the HR5 galaxy catalog as follows. We first select galaxies by imposing the stellar mass criterion. 
To have a good IA signal, we restrict our galaxy sample to have $M_{\star} \geq 10^{10}\ {\rm M}_{\odot}$, which corresponds to about 5000 stellar particles. The most massive subhalo sample tends to show the highest IA amplitude \citep{Tenneti2015a,Tenneti2015b,Chisari2015}.

As the simulation is running down to the final redshift, the low-level DM particles in the outside of the zoomed in region (therefore they have a larger mass) may infiltrate and contaminate the boundaries of the zoomed region more seriously. This may affect the kinematics of stellar particles of galaxies. To minimize this effect, we flag galaxies that are at least $2.052\ h^{-1}\ \mathrm{cMpc}$ away from low-level DM particles. This flag is used to remove contaminated galaxies, resulting in a clean sample consisting exclusively of level-13 particles, which is used in the subsequent analysis.  When a galaxy is excluded, we assign more weight to its neighboring galaxy to compensate for the exclusion.

Also, some ongoing mergers may not be relaxed, and early-stage mergers can therefore be misidentified during galaxy identification, introducing potential biases. To mitigate this effect, we use the asymmetry parameter for visual inspection and exclude unrelated systems.

To classify star-forming main-sequence and quiescent galaxies, the diagram of star-formation rate (SFR) and stellar mass has widely been applied  \citep{Daddi2007,Elbaz2007,Noeske2007}. This approach appears to be reliable for selecting early-type galaxies. However, some of early-type galaxies undergo active star formation, which is due to the external origin like galaxy mergers and gas accretion from the cosmic web \citep{Lee2023}. Also, some late-type galaxies have low star formation activities \citep{Paspaliaris2023}. Therefore, we choose to use another criterion based on kinematics. Because early-type galaxies are considered to be supported by the stellar velocity dispersion, we utilize the ratio ($|V_{\rm r}/\sigma_V|$) between the rotational velocity and the velocity dispersion to refine the classification of early-type galaxies. We note that our criterion does not incorporate Sérsic index for classifying galaxies based on their spatial morphology \citep{Park2022}.

\begin{figure}[!t]
\centering
\includegraphics[width=0.47\textwidth]{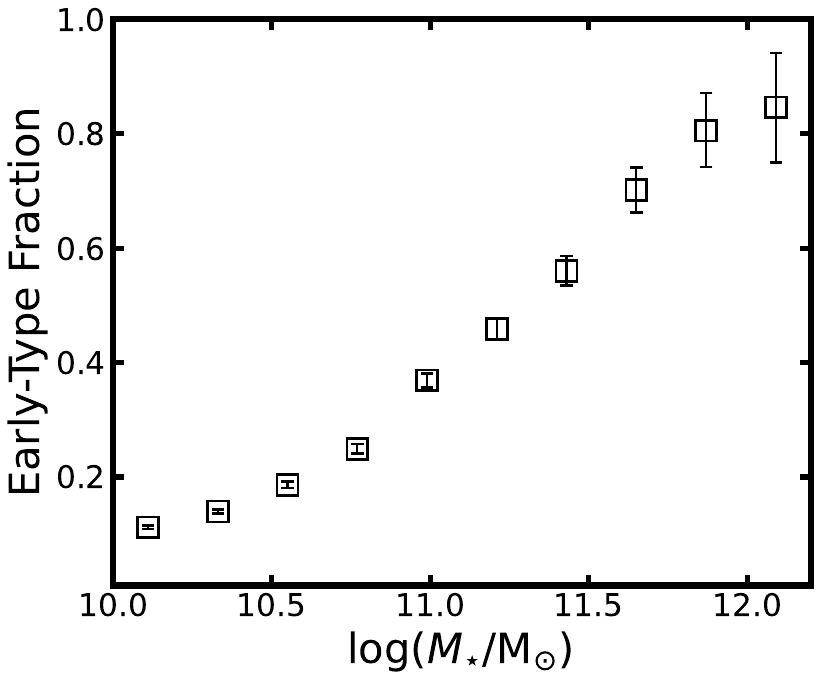}
\caption{A plot of fractions of early-type galaxies as a function of stellar mass. The fraction increases with stellar mass, a trend consistently observed across different redshifts. The error bars are obtained from bootstrap sampling.
\label{fig:redfraction}}
\end{figure}
Figure~\ref{fig:Mstar_SFR} shows the SFR-stellar mass diagram for HR5 galaxies after applying the mass threshold at $z=0.625$. We plot the SFR averaged over 100 Myr. 
We perform a least-square fit to identify the star-forming main sequence, and define quiescent galaxies as those lying in a factor of four below the main sequence to accommodate observational findings \citep[e.g.,][]{Kim2017,Lee2023}.

As mentioned earlier, some of early-type galaxies could experience star formation. Therefore, we may miss them according to the SFR criteria. Therefore, we supplement the method with the kinematic morphology, $|V_{\rm r}/\sigma_V|$.
The values of $V_{\rm r}$ and $\sigma_V$ of a galaxy are calculated from the stellar components within the radial range $R_{1/3} < r < R_{2/3}$ in the cylindrical coordinate by setting the direction of angular momentum as the $z$ axis, where $R_{1/3 (2/3)}$ represents the radius that encloses one-third (two-thirds) of the galaxy stellar mass. Figure~\ref{fig:SFR_ratio} displays star forming galaxies and quiescent galaxies on the plane of SFR versus $|V_{\rm r}/\sigma_V|$. Quiescent galaxies are distinct compared to star forming galaxies in the kinematic morphology. The majority of quiescent galaxies are distributed under $|V_{\rm r}/\sigma_V| = 0.5$, while star forming galaxies show skewed Gaussian distribution centered at $|V_{\rm r}/\sigma_V| = 0.79$. For early types, we select $|V_{\rm r}/\sigma_V| = 0.42$, which is identical to the median of the distribution of quiescent galaxies.

Figure~\ref{fig:target_selection} shows the final criteria used to select early-type galaxies at $z=0.625$, with $|V_{\rm r}/\sigma_V|$ plotted as a function of stellar mass. Galaxies satisfying the conditions for both stellar mass and kinematic ratio are highlighted in green, representing massive early-type galaxies. In Figure~\ref{fig:SFR_ratio}, we may confirm our sample selection using SFR. Figure~\ref{fig:redfraction} displays the fraction of early-type galaxies as a function of stellar mass, demonstrating that the fraction increases with stellar mass \citep{Nair2010,Vulcani2011,Pfeffer2023,Lee2024}. The same selection criteria are applied to all redshifts for a consistent analysis of early-type galaxies. 
Table~\ref{tab:sample_size} gives the sample size at different redshifts. We note that the numbers of early-type galaxies are different from those of hosting subhalos at some redshifts. In HR5, there are some galaxies with very few dark matter particles, which are called dark matter-deficient galaxies. The number of dark-matter particles of these galaxies is not enough to measure the shapes. Therefore, such objects are excluded when conducting analyses.
\begin{deluxetable}{lcccccc}[!h]
\tabletypesize{\scriptsize}
\tablewidth{1\textwidth} 
\renewcommand{\arraystretch}{1.5}
\tablecaption{The number of galaxies and hosting subhalos with $M_\star \geq 10^{10}\ \rm{M}_\odot$\label{tab:sample_size}}
\tablehead{
\colhead{$ z $} & & \colhead{Total Galaxy} & & \colhead{Early-Type} & & \colhead{Hosting Subhalo}
} 
\startdata 
 0.625   &  &   $37462$   &  &   $5335$   &  &   $5335$   \\ 
 1.0     &  &   $33673$   &  &   $4469$   &  &   $4467$   \\
 1.5     &  &   $26469$   &  &   $3562$   &  &   $3561$   \\
 2.0     &  &   $20023$   &  &   $2617$   &  &   $2613$   \\
 2.5     &  &   $13397$   &  &   $1365$   &  &   $1365$   
\enddata
\end{deluxetable}

\section{METHOD} \label{sec:method}

\subsection{Shape Measurement} \label{subsec:shape}

The accurate measurement of galaxy shapes is essential for the study of both the weak lensing and the IA. In observations, various techniques are used to measure galaxy shapes. One method involves applying model-fitting algorithms, such as Lensfit \citep{Miller2007,Miller2013} and Im3shape \citep{Zuntz2013} to the observed surface brightness distribution. Another approach is to directly measure the ellipticity using the quadrupole moments of the surface brightness \citep{Kirk2015}. 

In simulations, the inertia tensor is widely used to parameterize galaxy shapes \citep[e.g.,][]{Bailin2005b,Kiessling2015}. In this study, we implement this method to measure the projected shapes of galaxies and dark-matter subhalos. The reduced inertia tensor is computed to quantify shape parameters of galaxies and subhalos using the positions and masses of stellar  and dark-matter particles. It is defined as follows;
\begin{equation} \label{eq:inertia}
I_{ij} \equiv \frac{1}{M} \sum^n_{k=1}{m_k \frac{x_{k,i}x_{k,j}}{r^2_k}},
\end{equation}
where $m_k$ is the mass of $k$-th particle and $x_{k,i}$ is the $i$-th component of the ``projected'' position vector of the particle with respect to the center of mass. The term $r_k$ denotes the weighted distance from a particle to the center of the galaxy or dark-matter subhalos. 

The inertia tensor has two eigenvectors (${\bf e}_a$ and ${\bf e}_b$) and their corresponding eigenvalues ($\lambda_a$ and $\lambda_b$). The axis ratio of major and minor axes of the projected ellipse is obtained as $\sqrt{\lambda_a/\lambda_b}$. From now on, we assume $\lambda_a\ge \lambda_b$. In the process of shape measurement, we initially adopt the spherical weighting, 
\begin{equation}
    r^2_k = x^2_{k,i} + x^2_{k,j},
\end{equation}
and repeat applying the elliptical weighting in the iteration process,
\begin{equation}
   r^2_k = \left(\frac{\mathbf{x}_k \cdot \mathbf{e}_a}{a} \right)^2 + \left(\frac{\mathbf{x}_k \cdot \mathbf{e}_b}{b} \right)^2,
\end{equation}
where $a\equiv \sqrt{\lambda_a}$ and $b\equiv \sqrt{\lambda_b}$. 
We iteratively measure the shape, updating $r_k$ until the axis ratio, $q\equiv b/a$, converges within $0.1\%$. In this calculation, we simply assume the $x$--axis of the simulation coordinate as the line-of-sight direction. The inertia tensor is, then, computed using the projected positions of particles on the $y$--$z$ plane, and the ellipticity is calculated by treating the inertia tensor elements as quadrupole moments. 

We adopt the relation between the ellipticity and the position angle as follows;
\begin{equation} \label{eq:e_PA}
    (e_1,e_2) = \left(\frac{1-q}{1+q}\right)(\cos{2\theta},\sin{2\theta}),
\end{equation}
where $e_1$, $e_2$ represent real and imaginary components of the ellipticity and $\theta$ means the position angle of a galaxy. To minimize the influence of outer substructures on the shape measurements, we limit the member particles used in equation (\ref{eq:inertia}). For galaxies, the boundary is set at three times the 3D stellar half-mass radius, $R_{1/2}^{\star}$, ensuring that at least 80\% of the stellar particles are included. For dark-matter subhalos, it is set at 20 times $R_{1/2}^{\star}$ where we use at least 60\% of member dark-matter particles for the most of dark-matter subhalos. These limiting values are chosen to reduce contamination from clumpy structures of outer part of galaxies and subhalos while ensuring that a sufficient number of particles are included to maintain the accuracy of shape measurements.

\subsection{Two-Point Statistics} \label{subsec:twopoint}

We use two-point statistics to quantify the IA of galaxies. The modified Landy-Szalay estimator \citep{Landy1993} is used to measure the shape-density cross-correlation function \citep{Mandelbaum2006a,Joachimi2011,Singh2015}:
\begin{equation} \label{eq:twopoint3d}
    \xi_{g+}\left(\mathbf{r}\right) = \frac{S_+D(\mathbf{r}) - S_+R(\mathbf{r})}{RR(\mathbf{r})},
\end{equation}
where $\mathbf{r}$ represents the separation vector of a galaxy pair. The term $RR$ represents the pair count of randomly distributed points. The term $S_+D$ means the sum of the + component of the $j$-th galaxy that is measured with the $i$-th galaxy:
\begin{equation} \label{eq:SD}
    S_+D(\mathbf{r}) \equiv \sum_{i \ne j |\mathbf{r}}{w_j e_+\left(j|i\right)},
\end{equation}
where $w_j$ indicates the normalized weight factor of the $j$-th galaxy correcting for the exclusion of contaminated galaxies. The term $e_+\left(j|i\right)$ represents the relative ellipticity of the $j$-th galaxy viewed from the $i$-th galaxy in the sample,
\begin{equation} \label{eq:e+}
    e_+\left(j|i\right) \equiv e_1 \cos2\varphi + e_2 \sin2\varphi,
\end{equation}
where $e_1$ and $e_2$ indicate the ellipticity of the $j$-th galaxy, and $\varphi$ means the position angle between the line of galaxy pair and the reference axis of the projection plane ($y$--axis).

The covariance matrix of the two-point correlation function is computed using the jackknife resampling method. We define a ``clean'' region where the low-level contamination is minimized. The region has a volume of $(717\times65\times68) \left(h^{-1}\ \mathrm{cMpc}\right)^3$. For the jackknife resampling we divide the region into 80 sub-boxes of size $\left(35.85\times32.5\times34\right) \left(h^{-1}\ \mathrm{cMpc}\right)^3$, which are sufficient in number and volume to estimate the covariance matrix. 

\subsection{NLA Model} \label{subsec:nla_model}

The linear alignment (LA) model has been one of the most prevailing approaches to the IA of galaxy shapes \citep{Catelan2001,Hirata2004,Blazek2011}. In the model, the intrinsic shape of a galaxy is determined by the tidal field of the large-scale structure at the formation epoch of the galaxy. Instead of using the linear power spectrum as in the LA model, the non-linear alignment (NLA) model implements the non-linear power spectrum \citep{Bridle2007,Hirata2007} for a better description of the clustering of matter field. 

\begin{sidewaysfigure}
\centering
\includegraphics[width=1\textwidth]{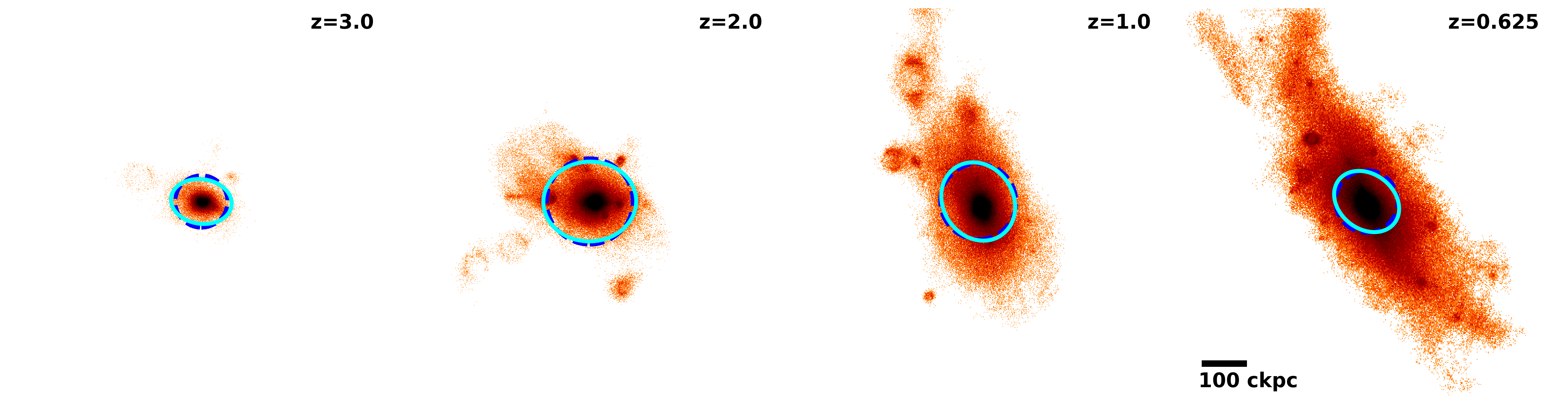}
\caption{A distribution of stellar particles of the most massive galaxy at $z=0.625$ with $M_\star = 1.436 \times 10^{12}\ {\rm M}_{\odot}$. Its progenitors are traced back at $z=1$, 2, and 3. The blue dashed circle represents the boundary of particle selection which has the value of $3R_{1/2}^{\star}$ for a galaxy. The shape of a galaxy is determined using its inertia tensor iteratively. We obtain the shape parameters such as ellipticity, major and minor axes, and position angle. The cyan solid ellipse illustrates the resulting shape.
\label{fig:galaxyshape}}
\end{sidewaysfigure}

\begin{sidewaysfigure}
\centering
\includegraphics[width=1\textwidth]{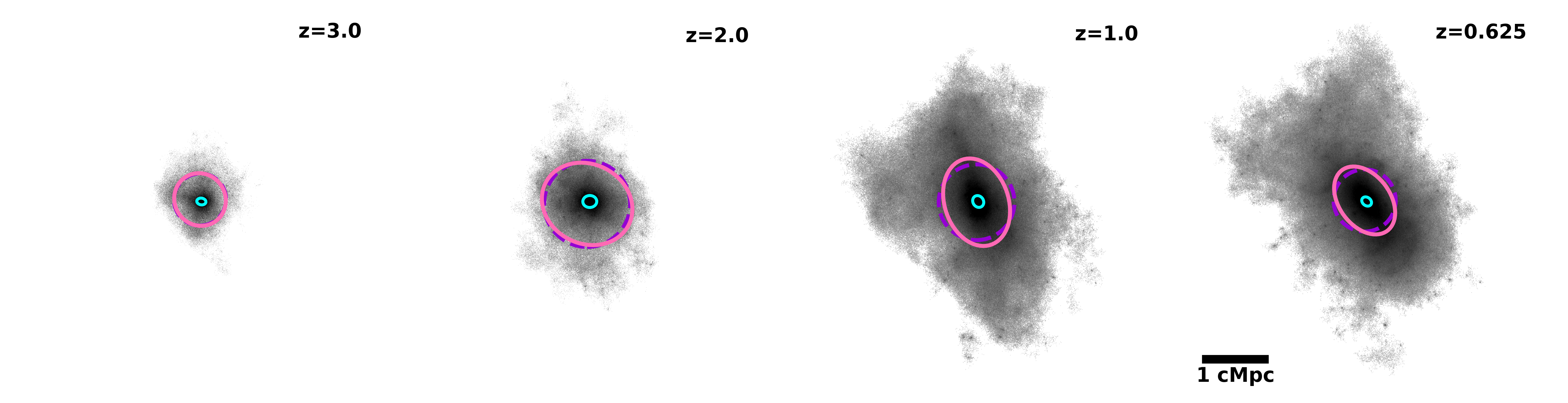}
\caption{Similar to Figure~\ref{fig:galaxyshape}, but for the hosting dark-matter subhalo. The boundary of particle selection for dark-matter subhalo is $20R_{1/2}^{\star}$ denoted as the dark-violet dashed circle. Resulting shape is shown with the bright-violet solid ellipse. We add the shape of galaxy with the cyan solid ellipse for comparison.
\label{fig:dmshape}}
\end{sidewaysfigure}

In the matter-dominated era, the intrinsic shape ($ e^{\rm I}_1,\,e^{\rm I}_2$) of a galaxy is related to the primordial gravitational potential ($\phi_{\rm p}$) in the following way;
\begin{equation} \label{eq:ellipticity}
    \left(e^{\rm I}_1,\,e^{\rm I}_2\right) = -\frac{C_1}{4\pi G}\left(\frac{\partial^2}{\partial x^2} - \frac{\partial^2}{\partial y^2} \,, 2\frac{\partial^2}{\partial x \partial y} \right)\phi_{\rm p},
\end{equation}
where $C_1$ is a normalization factor. The primordial potential is connected with the overdensity $\delta(\mathbf{k})$ through the Poisson equation in the Fourier space,
\begin{equation} \label{eq:poisson}
    \phi_{\rm p}(\mathbf{k})=-4 \pi G \frac{\bar{\rho}\left(z\right)}{\left(1+z\right)D\left(z\right)} a^2 k^{-2} \delta(\mathbf{k}),
\end{equation}
where $a$ means the cosmic scale factor, $\bar\rho(z)$ is the mean matter density at $z$, and $D(z)$ is the linear growth factor normalized as $D(z=0)=1$. 
According to \cite{Hirata2004}, the cross power spectrum between the shape and density has the form;
\begin{equation} \label{eq:shapepower}
    P_{\delta {\rm I}}\left(\mathbf{k},z\right) = -\frac{C_1\bar{\rho}\left(z\right)}{\left(1+z\right)D\left(z\right)}a^2P^{\rm nl}_\delta\left(\mathbf{k},z\right).
\end{equation}

The theoretical modeling of IA correlations is well studied in three dimensions \citep{Okumura2020a,Okumura2020b}. We adopt this formalism to convert the power spectrum ($P_{\delta {\rm I}}$) of \cite{Hirata2004} into the cross correlation of shape and density in three dimensions as follows;
\begin{equation} \label{eq:xi_gI}
    \xi_{g{\rm I}}\left(\mathbf{r},z\right) = (1-\mu^2)b_g\int^\infty_0\frac{k^2dk}{2\pi^2}P_{\delta {\rm I}}(k,z)j_2(kr),
\end{equation}
where $b_g$ means the galaxy bias, and $\mu$ represents the cosine between separation vector and line-of-sight direction indicating the angular dependence term $(1-\mu^2)$ corresponding to the shape projection onto a plane \citep{Okumura2019}. The term, $j_2(kr)$, means the second-order spherical Bessel function. We use the positions of galaxies as tracers of the density field. To account for the bias between galaxy density and matter density, we adopt the linear galaxy bias.

\section{RESULTS} \label{sec:result}

\subsection{Sample Analysis} \label{subsec:sample}

\begin{figure}[!t]
\centering
\includegraphics[width=0.47\textwidth]{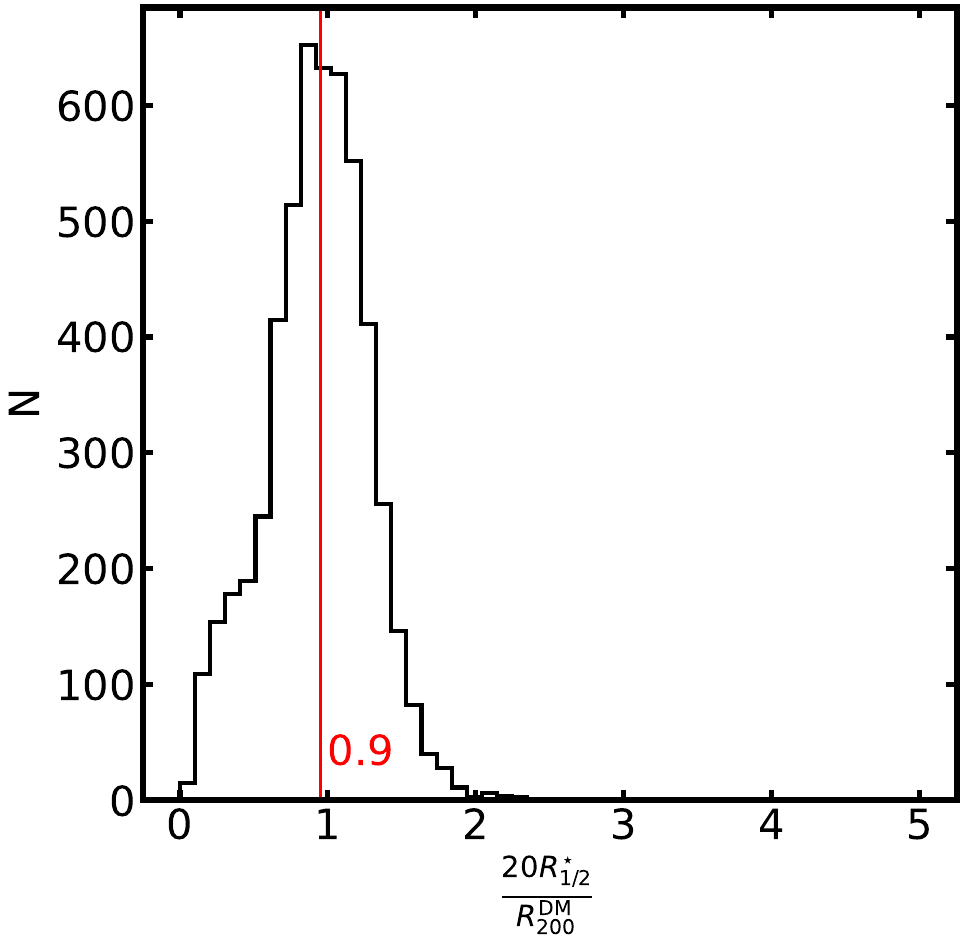}
\caption{A histogram of the ratio of $20R_{1/2}^{\star}$ and $R_{200}^{\rm{DM}}$. It is concentrated around unity, which demonstrates that our selection criterion (based on $R_{1/2}^{\star}$) is physically reasonable.
\label{fig:r200}}
\end{figure}

In this section, we present the statistical properties of the shapes of galaxies and dark-matter subhalos.  We illustrate the evolutionary phase of the progenitor of the most massive galaxy found at $z=0.625$ and its hosting dark-matter subhalo  in Figure ~\ref{fig:galaxyshape} and Figure \ref{fig:dmshape}, respectively. As described in Section~\ref{subsec:shape}, galaxy shapes are determined using the member stellar particles within three times the stellar half-mass radius, but the measurement for dark-matter subhalos extends to 20 times the stellar half-mass radius. The shapes are visually represented by solid ellipses in the figures. The oblateness and position angle of the measured shapes align well with the images of the galaxy and dark-matter subhalo, demonstrating the fairness of shape measurements.
\begin{figure}[!t]
\centering
\includegraphics[width=0.47\textwidth]{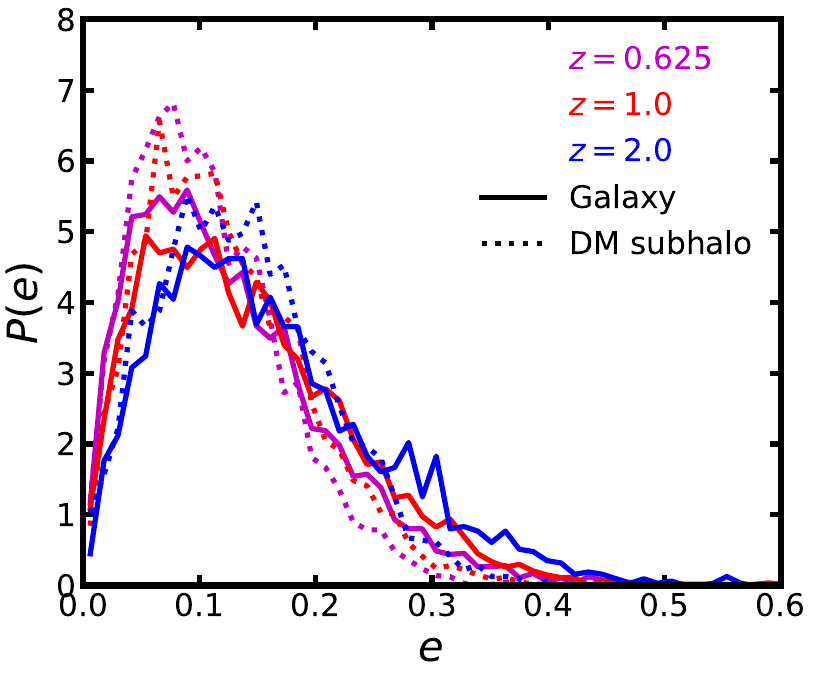}
\caption{Distributions of the total ellipticity, $e =\sqrt{e_1^2 + e_2^2}$, of galaxies (solid lines) and dark-matter subhalos (dotted lines) at $z=0.625$, 1.0, and 2.0. dark-matter shapes are rounder than galaxies. Both become rounder as they evolve with time.
\label{fig:edist}}
\end{figure}
\begin{figure}[!t]
\centering
\includegraphics[width=0.47\textwidth]{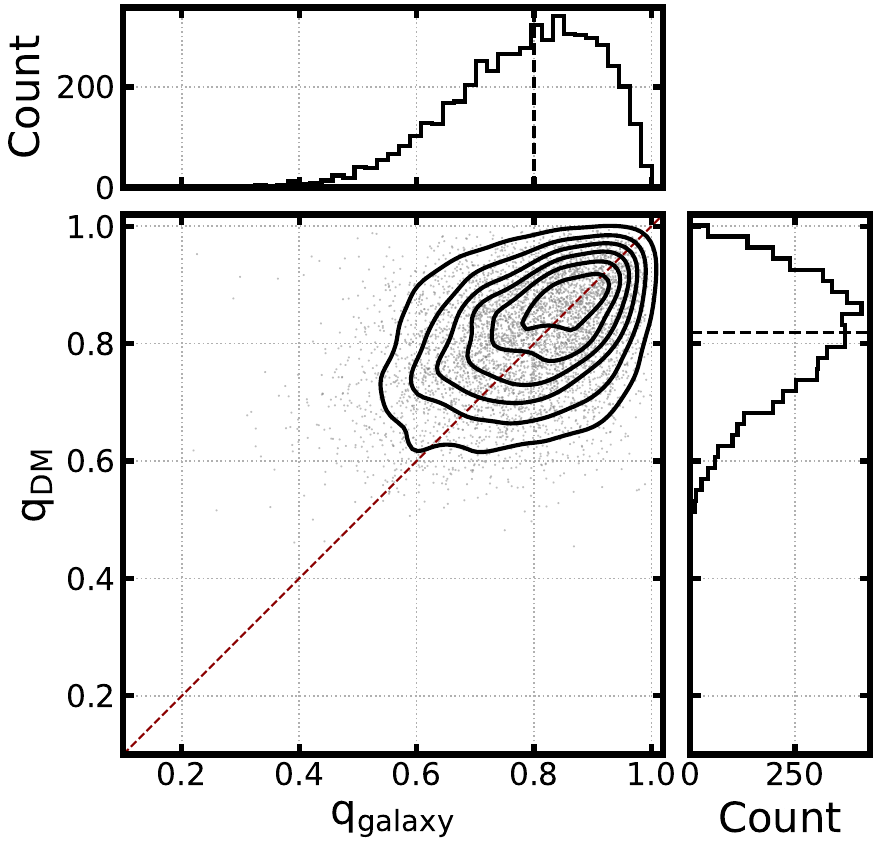}
\caption{A distribution of the axis ratio, $q=b/a$, of galaxies and dark-matter subhalos. Dashed lines in top and right panels represent the median of the distribution. The axis ratio for dark-matter subhalos, $q_{\rm DM}$, shows higher concentration than $q_{\rm galaxy}$ which indicates dark-matter subhalos are rounder than early-type galaxies. This feature is consistent with Figure~\ref{fig:edist}.
\label{fig:qvalue}}
\end{figure}
\begin{figure}[!t]
\centering
\includegraphics[width=0.47\textwidth]{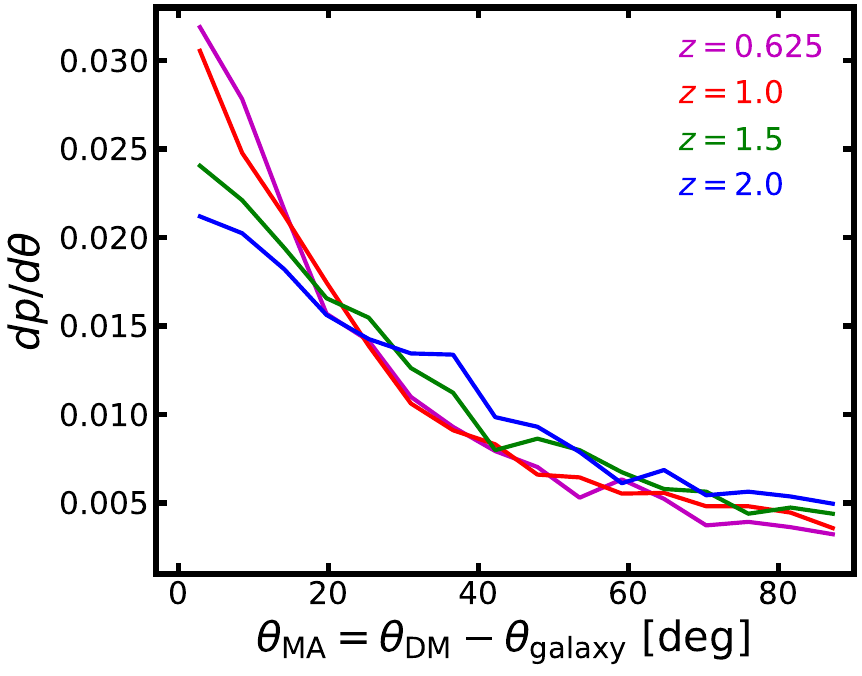}
\caption{Distributions of misalignment angle for 2D shapes of galaxies and dark-matter subhalos at $z=0.625$, 1.0, 1.5, and 2.0. Galaxies are more aligned with dark-matter subhalos at lower redshift. About 22\% of galaxies are misaligned with their dark-matter subhalos by a difference angle greater than $\theta_{\rm MA}=45^\circ$ at $z=0.625$. 
\label{fig:misalignment}}
\end{figure}

Rather than directly using the half-mass radius of dark matter, we use the half-mass radius of stellar matter for the subhalo analysis in order to reconcile with  observations, which mostly use the half-light radius (which is converted to half-mass radius with a proper mass-to-light ratio). Figure~\ref{fig:r200} shows  a distribution of the ratio between $20R_{1/2}^{\star}$ and $R_{200}^{\rm{DM}}$ for dark-matter subhalos. We note that for 82\% of subhalos, more than 80\% of member particles are used for shape measurements. Figure~\ref{fig:r200} indicates that 20 times of $R_{1/2}^{\star}$ closely approximates to $R_{200}^{\rm{DM}}$ for most of subhalos. The choice of aperture is critical, as the subhalo shape is highly sensitive to the extent of the boundary. A smaller boundary significantly weakens the IA signal, rendering it less prominent than the case of galaxies. It is well established that dark-matter subhalos exhibit higher amplitudes of shape-density correlations than galaxies \citep{Okumura2009b,Okumura2009,Tenneti2015b}. Our choice of $20R_{1/2}^{\star}$ reproduces the similar behavior (Section~\ref{subsec:tpcf}).
\begin{figure}[!t]
\centering
\includegraphics[width=0.47\textwidth]{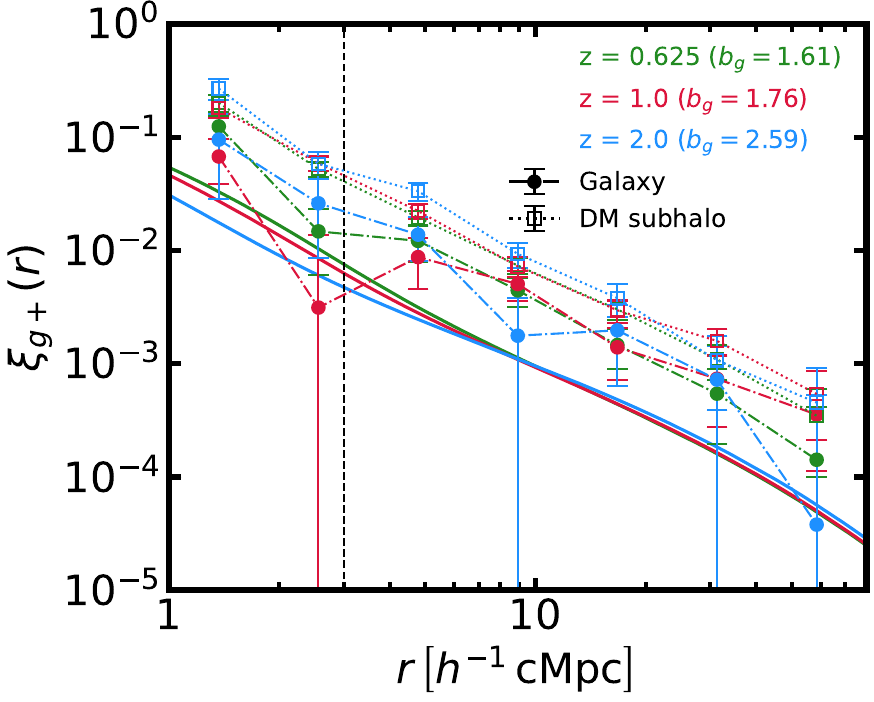}
\caption{Cross correlations, $\xi_{g+}\left(r\right)$, measured for early-type galaxies (filled circles) and dark-matter subhalos (open squares) at redshifts $z=0.625$, 1.0, and 2.0. Dark-matter subhalos show higher correlations than galaxies. The error bars represent the square root of the diagonal components of the covariance matrix. The solid lines show the theoretical predictions which are normalized by $C_{\rm{HS}}$ (see Section~\ref{subsec:IAamp}). The vertical dashed line marks $r = 3\, h^{-1}\ \rm{cMpc}$, above which scales are used to obtain the amplitude parameter. \label{fig:3dsignal}}
\end{figure}

We analyze ellipticity distributions of galaxies and dark-matter subhalos in Figure~\ref{fig:edist}. 
We find that galaxies and dark-matter subhalos become more spherical with time. We analyze the effect of various shape measurements on our results in Appendix~\ref{sec:app_a}. Figure~\ref{fig:qvalue} compares the distribution of axis ratio for galaxies and dark-matter subhalos. These results indicate that the shape of a galaxy tends to be more elliptical than the hosting subhalo \citep{Pulsoni2021}. We note that the shape measurements for galaxies are relatively robust because of a large number of particles involved, as we focus on massive galaxies with $M_\star \geq 10^{10}\ \rm{M}_\odot$. In contrast, some subhalos have member particles fewer than 100, which can introduce a significant noise in the shape measurement. Figure~\ref{fig:misalignment} demonstrates that HR5 galaxies and subhalos experience some misalignment in the position angle as found in previous studies \citep{Okumura2009,Tenneti2014,Bhowmick2020}. We find that shapes of early-type galaxies become more aligned with surrounding dark-matter subhalos with time. 

\subsection{The Two-Point Correlation} \label{subsec:tpcf}

We examine the cross-correlation function, $\xi_{g+}(r)$, between intrinsic shapes of galaxies/subhalos and their positions at $z=0.625$, 1.0, and 2.0. This quantity is free from the grid locking effect that occurs in the simulations using the AMR scheme based on the octree mesh in Cartesian coordinates \citep{Chisari2015}. We verify this issue in Appendix~\ref{sec:app_b}.
\begin{figure}[!t]
\centering
\includegraphics[width=0.47\textwidth]{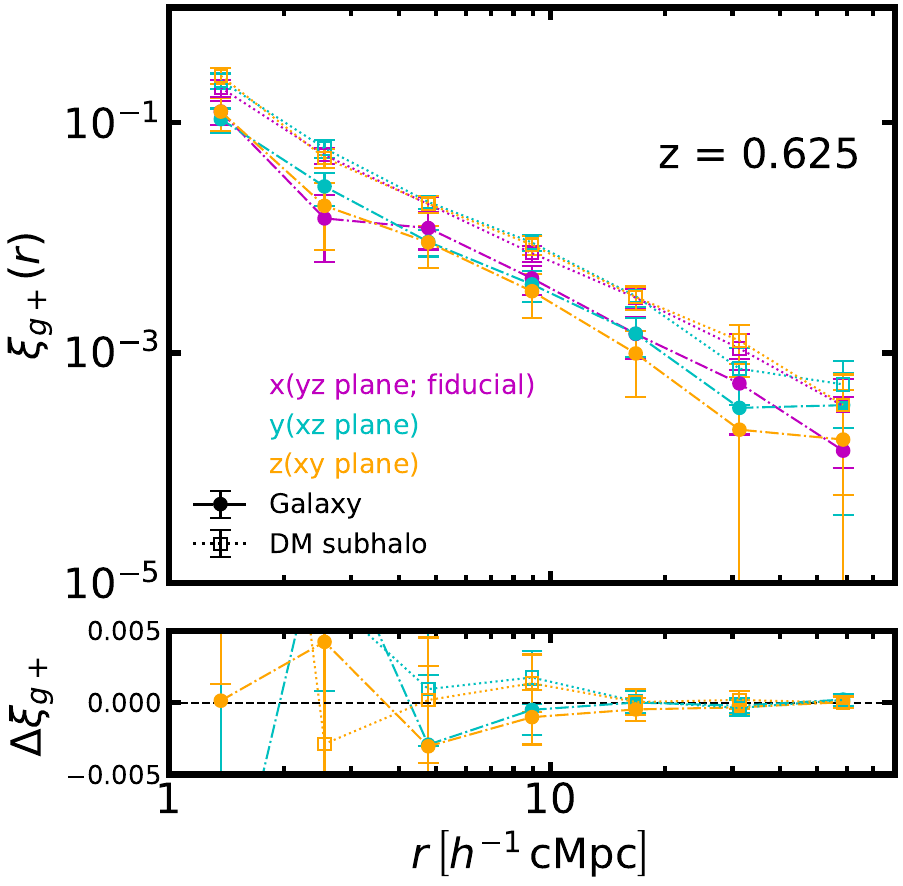}
\caption{Similar to Figure~\ref{fig:3dsignal} but with different directions of projection. ({\it top panel}): The shape-density correlations are measured at $z=0.625$. ({\it bottom}): The differences of the correlations from the fiducial ($x$-) axis are shown. All three correlations are consistent with each other, indicating that the projection axis does not make the systematics.
\label{fig:projection3d}}
\end{figure}

The $x$--axis of the simulation box is designated as the line-of-sight direction along which projected shapes are measured. The pair separation of interest has a range of $r = 1 $ -- $ 80\, h^{-1}\ \rm{cMpc}$. The lower limit is set to exclude the one-halo dominant regime \citep{Schneider2010,Tenneti2015b} and upper limit ensures that measurements are made in the clean region of HR5. Figure~\ref{fig:3dsignal} shows the cross correlation functions for galaxies and subhalos. The error bars show the square root of diagonal components of the covariance matrix. The IA signal of subhalos is stronger than galaxies. The cross correlation ($\xi_{g+}$) do not show a significant redshift dependence in our analysis. However, the quantity shown in Figure~\ref{fig:3dsignal} is proportional not only to the amplitude of IA, but also to the galaxy bias ($b_g$). We will discuss this further in detail in Section~\ref{sec:discussion}.
\begin{figure}[!t]
\centering
\includegraphics[width=0.47\textwidth]{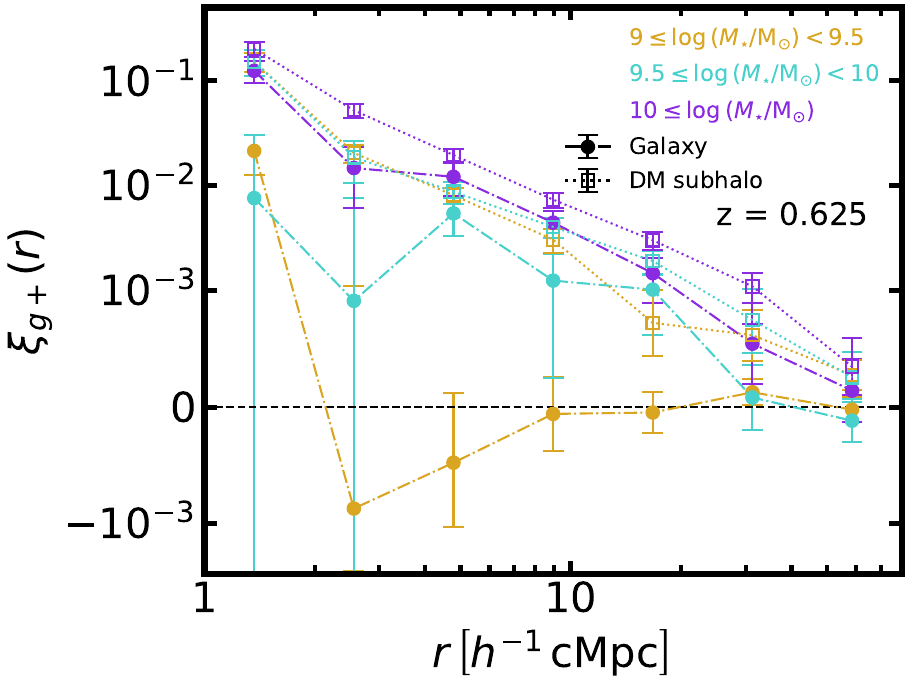}
\caption{Similar to Figure~\ref{fig:3dsignal} but with different stellar mass bins, $10^{9-9.5}\ \rm{M}_{\odot}$, $10^{9.5-10}\ \rm{M}_{\odot}$, and $\geq 10^{10}\ \rm{M}_{\odot}$. For galaxies, the correlation of the first bin is consistent with zero.
\label{fig:massdep}}
\end{figure}
\begin{figure}[!t]
\centering
\includegraphics[width=0.47\textwidth]{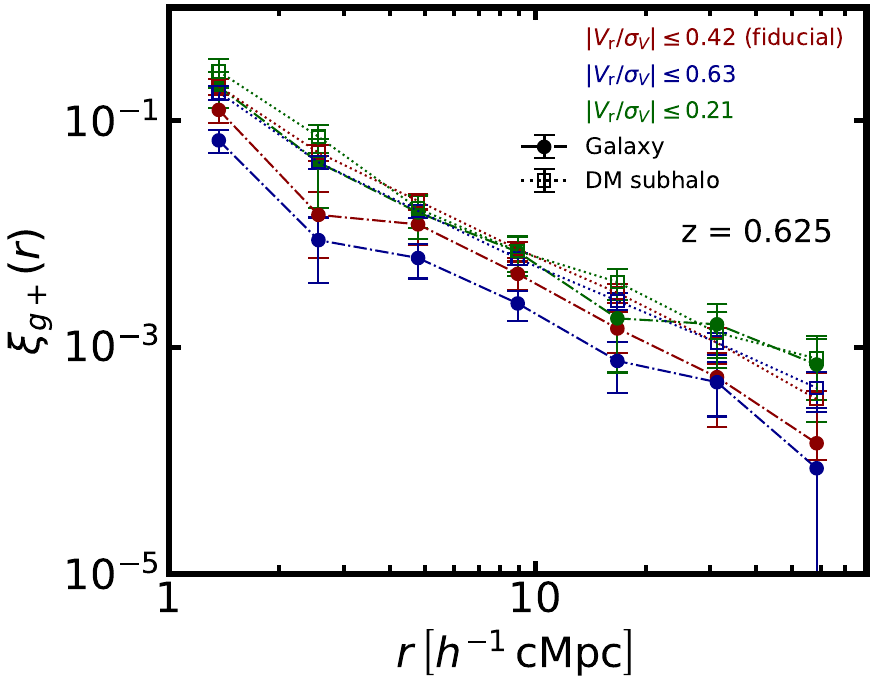}
\caption{Similar to Figure~\ref{fig:3dsignal} but with different kinematic morphology criterion. For galaxies, the correlation depends on the fraction of late-type galaxies. Dark-matter subhalos do not show the sensitivity with $|V_{\rm r}/\sigma_V|$.
\label{fig:ratiodep}}
\end{figure}
\begin{figure}[!t]
\centering
\includegraphics[width=0.47\textwidth]{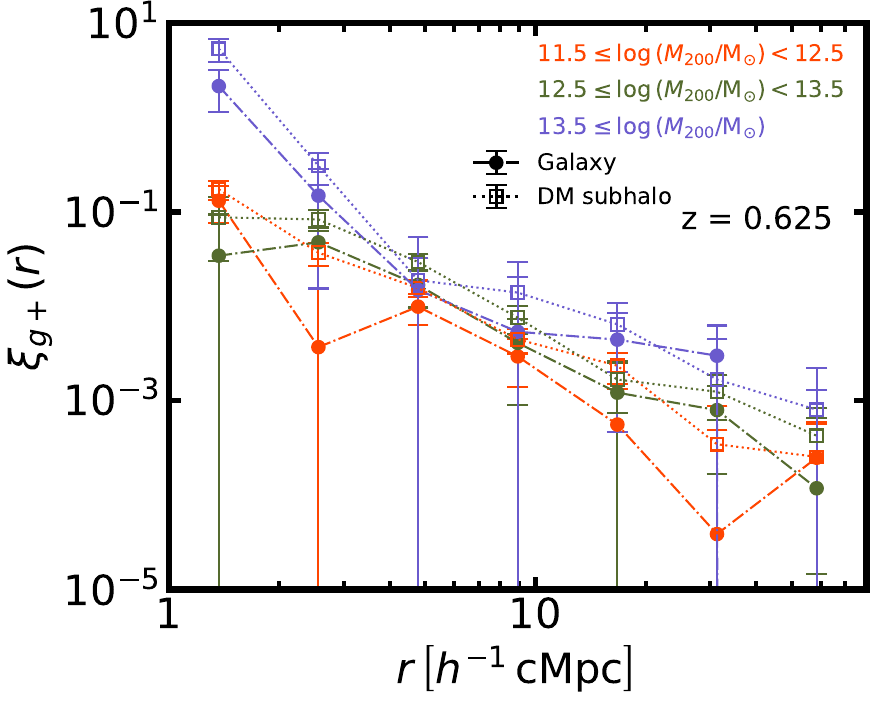}
\caption{Similar to Figure~\ref{fig:3dsignal} but with different halo mass bins, $10^{11.5-12.5}\ \rm{M}_{\odot}$, $10^{12.5-13.5}\ \rm{M}_{\odot}$, and $\geq 10^{13.5}\ \rm{M}_{\odot}$.
\label{fig:halodep}}
\end{figure}
To assess potential systematic biases related to the choice of the projection axis, we measure the shape along different line-of-sight directions and calculate the IA correlation at $z=0.625$, which is displayed in Figure~\ref{fig:projection3d}. The bottom panel shows the differences between the correlations for alternate projection planes and the fiducial projection direction, $x$--axis. 
Our analysis confirms that the choice of the projection direction does not introduce a significant bias into the results.

Figure~\ref{fig:massdep} shows the dependence of IA on stellar mass. For this plot, the stellar mass is divided into three bins, $M_{\star}=10^{9-9.5}\ {\rm M}_{\odot}$, $M_{\star}=10^{9.5-10}\ {\rm M}_{\odot}$, and $M_{\star}\geq10^{10}\ {\rm M}_{\odot}$. The same galaxy selection criteria and the measurement range are applied as used in Figure~\ref{fig:3dsignal}. Our results are consistent with previous findings \citep{Tenneti2015b} showing that the IA signal gets stronger with increasing stellar mass. For galaxies in the lowest mass bin, the IA signal is consistent with zero indicating no significant alignment. This feature agrees with Figure~\ref{fig:redfraction} because the fraction of early-type galaxies increases with the stellar mass. In contrast, subhalos exhibit positive IA correlations across all mass bins showing a stronger alignment signal even at lower stellar masses of galaxies hosted by the subhalos.

Figure~\ref{fig:ratiodep} illustrates the effect on $\xi_{g+}$ of rotation-supported galaxies. The fiducial selection criterion for early-type galaxies is $|V_{\rm r}/\sigma_V| \leq 0.42$. We find that galaxies with smaller $|V_{\rm r}/\sigma_V|$ show a higher IA correlation amplitude. The error bars for the low-ratio sample ($|V_{\rm r}/\sigma_V| = 0.21$) are substantially larger than the other two samples. This is attributed to the smaller sample size, which limits the statistical robustness of the measurements for this subset. In contrast, the IA correlation for dark-matter subhalos shows no significant variation across different $|V_{\rm r}/\sigma_V|$ samples. This is because, even for rotating galaxies, the corresponding dark-matter subhalos have little rotational component, making them susceptible to the tidal stretching effect. 

We also investigate host halo mass dependence of IA correlation in Figure~\ref{fig:halodep}. Galaxies and subhalos are selected following the previously described criteria and categorized based on the virial mass of their host halo $(M_{200})$ into three mass bins: $M_{200}=10^{11.5-12.5}\ {\rm M}_{\odot}$, $M_{200}=10^{12.5-13.5}\ {\rm M}_{\odot}$, and $M_{200}\geq10^{13.5}\ {\rm M}_{\odot}$.
Both galaxies and subhalos show stronger alignment signals with increasing host halo mass. Given the scatter in the stellar-to-halo mass relation, occasional overlaps and inversions in the measured signals at certain points happen.

\subsection{IA Amplitude Parameter} \label{subsec:IAamp}

In this subsection, we examine the evolution of the IA strengths for both the galaxies and the subhalos as a function of redshift. As described in Section~\ref{subsec:nla_model}, the parameter, $C_1$, which denotes the amplitude of the IA power spectrum, is normalized to a dimensionless parameter, $A_{\rm NLA}$, as follows;
\begin{equation} 
    A_{\rm NLA} = \frac{C_1}{C_{\rm{HS}}},
\end{equation}
where $ C_{\rm HS} =  5 \times 10^{-14}\ \rm{M}^{-1}_{\odot}\ h^{-2}\ \mathrm{Mpc}^3$ \citep{Brown2002,Hirata2004}, which is obtained by comparison with SuperCOSMOS \citep{Hambly2001}. The value of $A_{\rm NLA}$ is determined by comparing the measured shape-density correlation to the corresponding theoretical model. The fitting is performed over the range $3 < r < 80\, h^{-1}\ \rm{cMpc}$. Because galaxy positions are treated as biased tracers of the underlying density field, the galaxy bias $b_g$ is needed for theoretical predictions of $\xi_{g+}$ (equation~(\ref{eq:xi_gI})). We follow \cite{Einasto2023} to measure $b_g$ as the square root of the ratio between the galaxy and dark-matter density-density correlation functions at $6.84\ h^{-1}\ \mathrm{cMpc}$. We apply the same $b_g$ for both galaxies and subhalos because each galaxy is hosted by a subhalo. The optimal value of $A_{\rm NLA}$ and its associated uncertainty are obtained through a $\chi^2$ minimization process performed over a range of redshifts.

\begin{figure}[!t]
\centering
\includegraphics[width=0.47\textwidth]{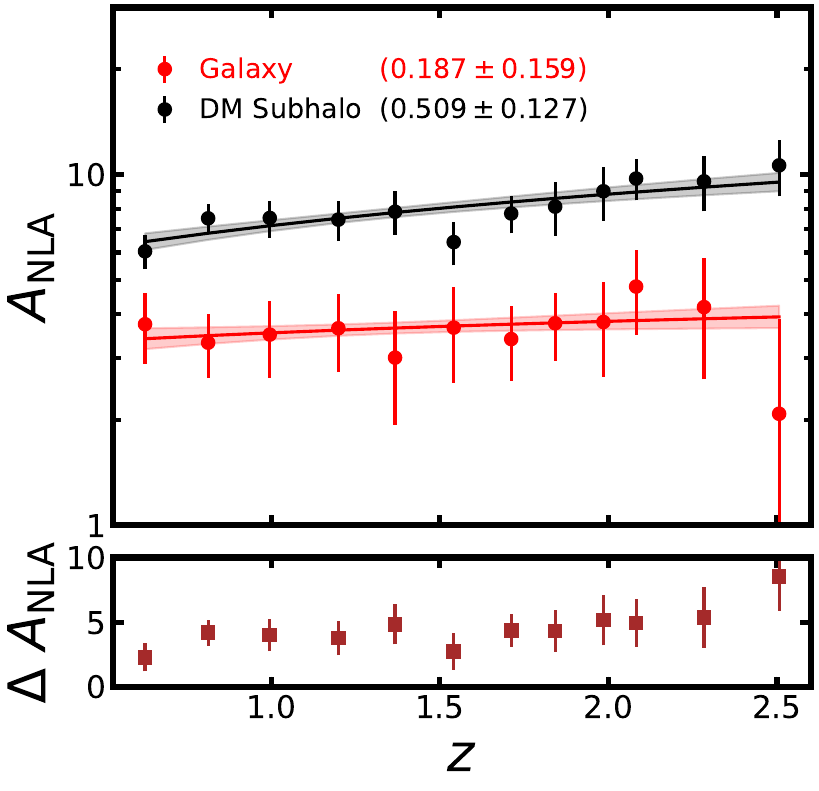}
\caption{IA amplitude parameter, $A_{\rm{NLA}}$, as a function of redshift. Red and black dots represent the $A_{\rm{NLA}}$ of galaxies and dark-matter subhalos, respectively. The range of redshift is from 0.6 to 2.5. Galaxies do not show evolutionary feature. However, dark-matter subhalos show evolutionary trend across redshifts. It is especially distinct at $z>1.5$. the bottom panel shows the difference of $A_{\rm{NLA}}$ between galaxy and dark-matter subhalo, $\Delta A_{\rm{NLA}} = A_{\rm{NLA}}^{\rm{DM}} - A_{\rm{NLA}}^{\rm{Galaxy}}$. \label{fig:amp3d}}
\end{figure}
Figure~\ref{fig:amp3d} shows the amplitude of the IA ($A_{\rm NLA}$) as a function of redshift. The evaluation of $A_{\rm NLA}$ stops at $z = 2.5$ because of the small sample size dropping below 1000 at higher redshifts. Subhalos always exhibit higher $A_{\rm NLA}$ than galaxies, which is consistent with the results presented in Section~\ref{subsec:tpcf}. While galaxies show no significant variation across redshifts, subhalos tend to show an increase in $A_{\rm NLA}$. We perform a weighted least-square fitting and obtain the power-law slope of $0.187 \pm 0.159$ for galaxies and $0.509 \pm 0.127$ for subhalos. We also find that the difference of $A_{\rm{NLA}}$ between galaxies and subhalos decreases at lower redshifts. This trend is consistent with the decreasing misalignments between galaxies and subhalos as shown in Figure~\ref{fig:misalignment}. The value of each data point in Figure~\ref{fig:amp3d} is listed in Table~\ref{tab:A_NLA_HR5}.
\begin{deluxetable}{lcccccc}[!h]
\tabletypesize{\scriptsize}
\tablewidth{1\textwidth} 
\renewcommand{\arraystretch}{1.5}
\tablecaption{$A_{\rm{NLA}}$ for galaxy and dark-matter subhalo with $M_\star \geq 10^{10}\ \rm{M}_\odot$ \label{tab:A_NLA_HR5}}
\tablehead{
\colhead{$ z $} & & & \colhead{Galaxy} & & & \colhead{DM Subhalo}
} 
\startdata 
 0.625   &  &  &   $3.75^{+0.86}_{-0.86}$   &  &  &   $6.05^{+0.67}_{-0.67}$   \\ 
 0.8     &  &  &   $3.32^{+0.69}_{-0.69}$   &  &  &   $7.51^{+0.74}_{-0.74}$   \\
 1.0     &  &  &   $3.50^{+0.86}_{-0.86}$   &  &  &   $7.53^{+0.91}_{-0.91}$   \\
 1.2     &  &  &   $3.65^{+0.91}_{-0.91}$   &  &  &   $7.45^{+0.98}_{-0.98}$   \\ 
 1.36    &  &  &   $3.01^{+1.07}_{-1.07}$   &  &  &   $7.84^{+1.11}_{-1.11}$   \\
 1.5     &  &  &   $3.66^{+1.11}_{-1.11}$   &  &  &   $6.42^{+0.91}_{-0.91}$   \\
 1.7     &  &  &   $3.40^{+0.82}_{-0.82}$   &  &  &   $7.75^{+0.95}_{-0.95}$   \\ 
 1.8     &  &  &   $3.77^{+0.83}_{-0.83}$   &  &  &   $8.11^{+1.40}_{-1.40}$   \\
 2.0     &  &  &   $3.80^{+1.15}_{-1.15}$   &  &  &   $8.97^{+1.58}_{-1.58}$   \\
 2.1     &  &  &   $4.80^{+1.31}_{-1.31}$   &  &  &   $9.75^{+1.30}_{-1.30}$   \\ 
 2.3     &  &  &   $4.19^{+1.58}_{-1.58}$   &  &  &   $9.57^{+1.72}_{-1.72}$   \\
 2.5     &  &  &   $2.08^{+1.79}_{-1.79}$   &  &  &   $10.63^{+1.96}_{-1.96}$  \\
\enddata
\end{deluxetable}

\begin{figure}[!t]
\centering
\includegraphics[width=0.47\textwidth]{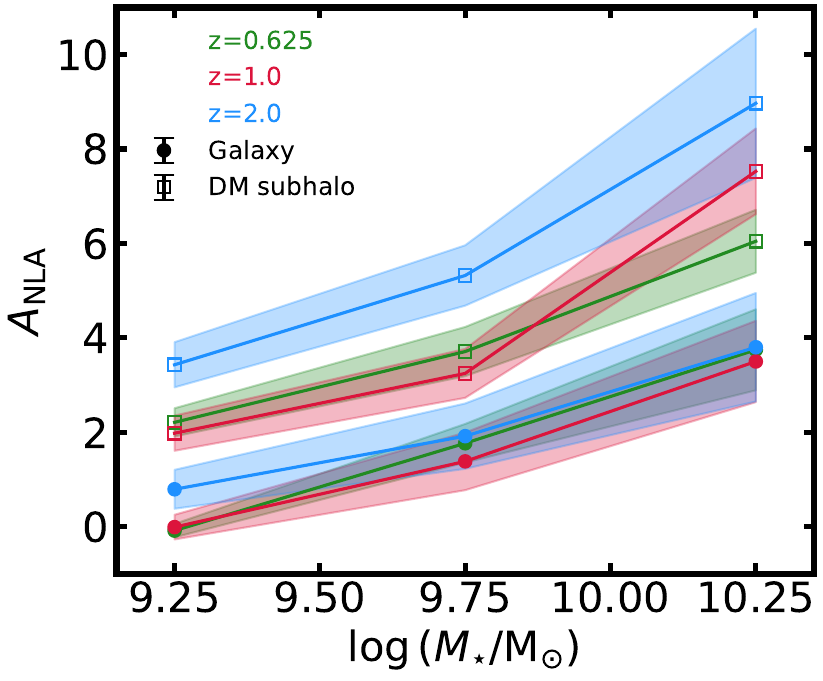}
\caption{The stellar mass dependence of $A_{\rm{NLA}}$. The mass bins are identical to Figure~\ref{fig:massdep}. As expected, $A_{\rm{NLA}}$ tends to increase as mass increases.
\label{fig:amp_mass}}
\end{figure}
\begin{figure}[!t]
\centering
\includegraphics[width=0.47\textwidth]{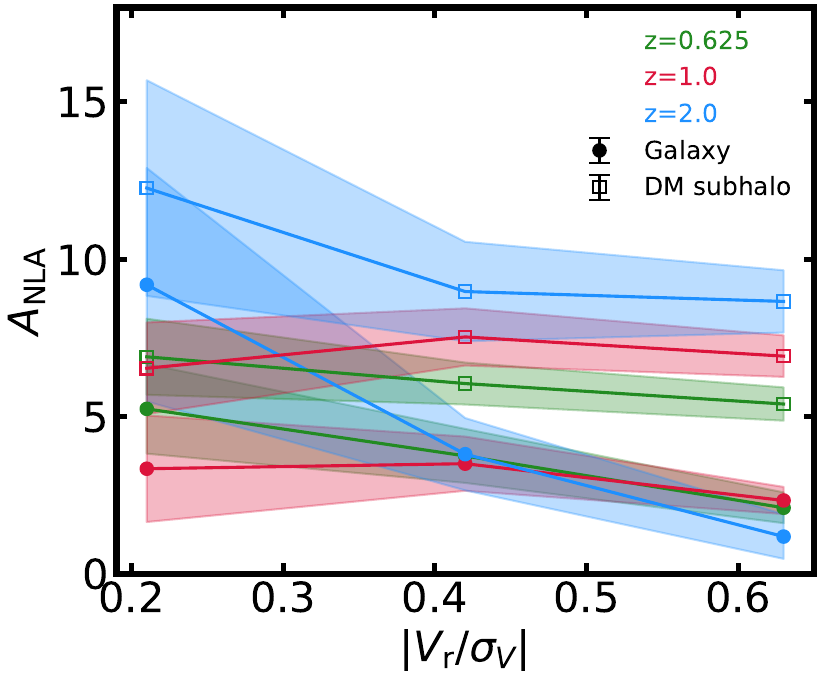}
\caption{Similar to Figure~\ref{fig:amp_mass}, kinematic morphology dependence of $A_{\rm{NLA}}$ are shown. As the ratio increases, $A_{\rm{NLA}}$ for galaxies decrease due to the effect of disk galaxies, whereas for dark-matter subhalos do not. 
\label{fig:amp_ratio}}
\end{figure}
\begin{figure}[!t]
\centering
\includegraphics[width=0.47\textwidth]{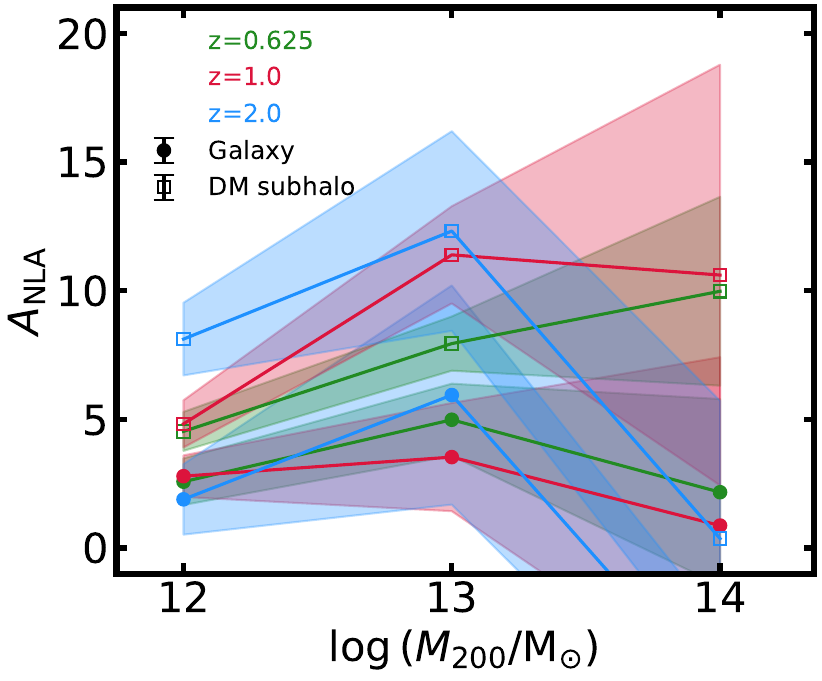}
\caption{Similar to Figure~\ref{fig:amp_mass}, but halo mass dependence of $A_{\rm{NLA}}$ are shown. The $x$--axis represents the median of each halo mass bin.
\label{fig:amp_Mhalo}}
\end{figure}
\begin{figure*}[!t]
\centering
\includegraphics[width=1.\textwidth]{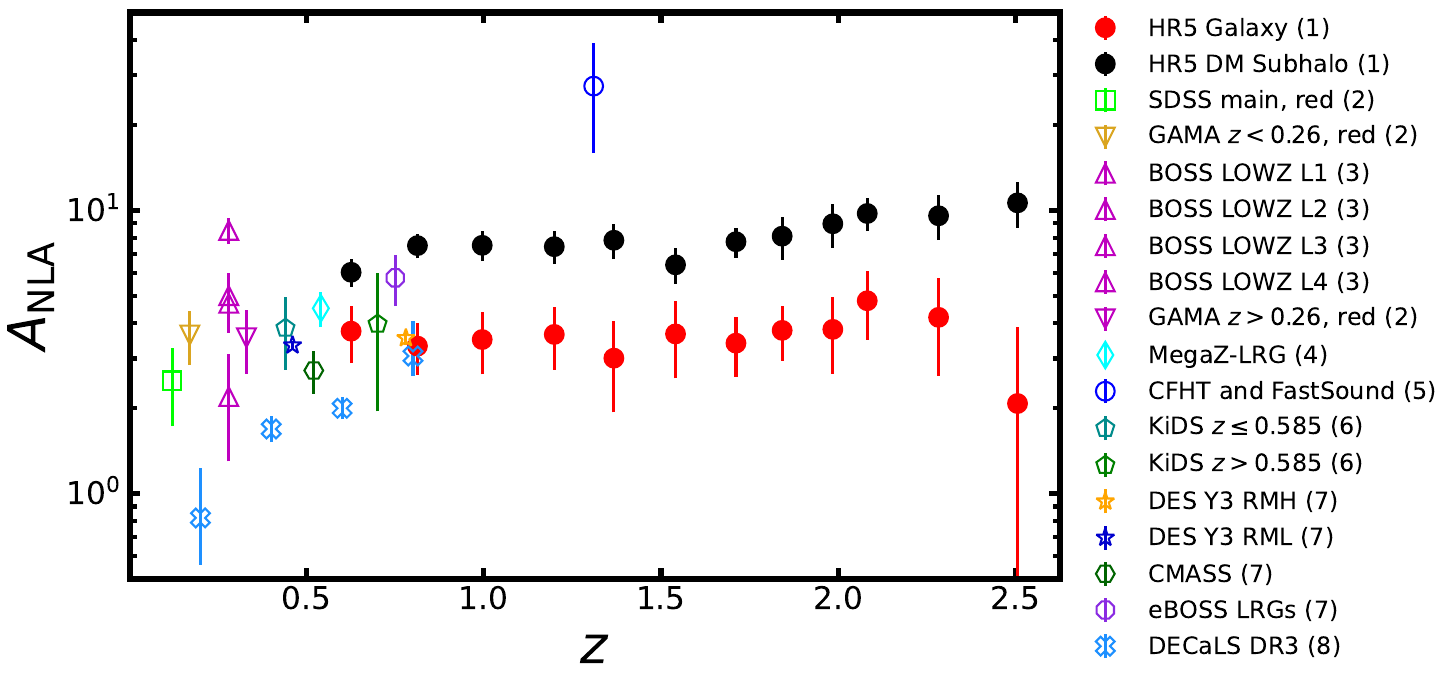}
\caption{Similar to Figure~\ref{fig:amp3d}, but with the $y$--axis on a log scale. The open symbols represent observational results. Our results are well matched with observations. We note that \cite{Yao2020}'s result overlaps with our results as well as other observations except for the first bin. Also \cite{Tonegawa2022}'s result deviates from our results. \textbf{References:} (1) This work; (2) \cite{Johnston2019}; (3) \cite{Singh2015}; (4) \cite{Joachimi2011}; (5) \cite{Tonegawa2022}; (6) \cite{Fortuna2021}; (7) \cite{Samuroff2023}; (8) \cite{Yao2020} \label{fig:amp3dobs}}
\end{figure*}
Figures~\ref{fig:amp_mass} and \ref{fig:amp_ratio} show the dependence of $A_{\rm NLA}$ on stellar mass and $|V_{\rm r}/\sigma_V|$, respectively. The amplitudes are measured for the three mass bins as described earlier. As in Figure~\ref{fig:massdep}, the amplitude increases with the inclusion of more massive galaxies at all redshifts. This trend, consistent with the results in Section~\ref{subsec:tpcf}, confirms that IA becomes stronger for higher-mass galaxies. In Figure~\ref{fig:amp_mass}, galaxies do not show a variation of the $A_{\rm NLA}$ for all mass bins and redshifts. For dark-matter subhalos, a nonzero amplitude is observed even in the lowest mass bin.  The amplitudes of dark-matter subhalo at $z=2.0$ are larger than lower redshift at all three mass bins. 

Figure~\ref{fig:amp_ratio} shows a decline in $A_{\rm NLA}$ of galaxies across all redshift ranges as $|V_{\rm r}/\sigma_V|$ increases, a trend that is consistent with the higher fraction of disk galaxies (blue) at higher $|V_{\rm r}/\sigma_V|$. If we use a criterion of higher $|V_{\rm r}/\sigma_V|$, the galaxy sample size is increased. However, it dilutes the alignment signal because of larger fraction of late-type galaxies in the sample. For dark-matter subhalos, this diminishing behavior is not obviously observed even with increasing $|V_{\rm r}/\sigma_V|$ cuts. 

We also investigate the halo mass dependence of $A_{\rm NLA}$ in Fig~\ref{fig:amp_Mhalo}. As halo mass increases from $10^{12}\ \rm{M}_{\odot}$ to $10^{13}\ \rm{M}_{\odot}$, the amplitude correspondingly increases. While subhalos exhibit a consistent increase in amplitude with redshift, galaxies show negligible evolution with redshift and display only a weak dependence on halo mass. In the highest mass bin, both galaxies and subhalos show decreased amplitude accompanied by significant uncertainties. These features result from limited statistical samples available at higher halo masses and redshifts.

\section{DISCUSSION} \label{sec:discussion}

\subsection{Comparison with Observations} \label{subsec:discuss1}

Figure~\ref{fig:amp3dobs} shows our results with observations, which are represented by open symbols with different colors. Our findings match most of observational results. Some observations, such as those by \cite[blue open crosses]{Yao2020} and \cite[dark blue open circles]{Tonegawa2022}, show trends inconsistent with the majority, which might be interpreted as indicative of redshift evolution. However, the amplitude parameter strongly depends on galaxy properties such as stellar mass, luminosity, and color, suggesting that different samples of galaxies can lead to large variation of the IA.

\begin{figure}[!t]
\centering
\includegraphics[width=0.47\textwidth]{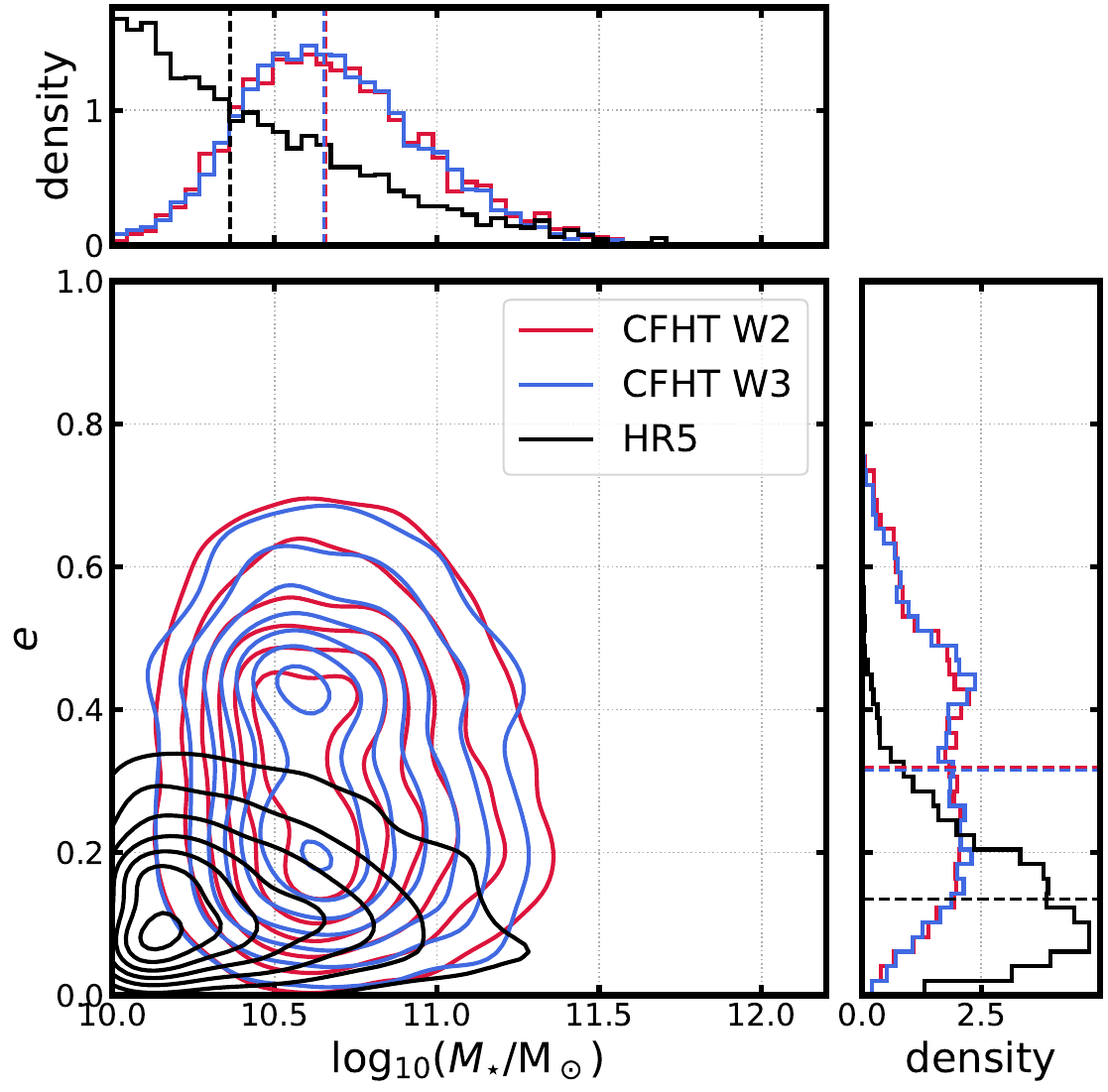}
\caption{Relations between stellar mass and measured ellipticity. Black lines represent HR5 galaxies. Red and blue lines represent CFHT Wide W2 and W3 fields, respectively. Top panel shows the stellar mass histogram. CFHT galaxies contains massisve galaxies than HR5 simulation. Right panel shows the distributions of ellipticity, similar to Figure~\ref{fig:edist}. We note that ellipticity of CFHT galaxies differ from HR5 galaxies showing much elliptical shapes.
\label{fig:Mstar_ellipticity}}
\end{figure}

To further investigate these differences, we examine the sample properties. Figure~\ref{fig:Mstar_ellipticity} displays the distributions of stellar mass and ellipticity for HR5 galaxies compared with the CFHTLenS sample used in \cite{Tonegawa2022} within the photometric redshift range of $1.13$ to $1.63$. The stellar mass distribution of HR5 differs significantly from that of CFHTLenS, which may explain why the correlation strength is higher than our results. Moreover, the ellipticity distributions of the HR5 galaxies and the CFHTLenS galaxies are notably different. These differences suggest that the observed trend arises from sample variance rather than redshift evolution.

\subsection{Comparison with Other Simulations} \label{subsec:discuss2}

\begin{figure}[!t]
\centering
\includegraphics[width=0.47\textwidth]{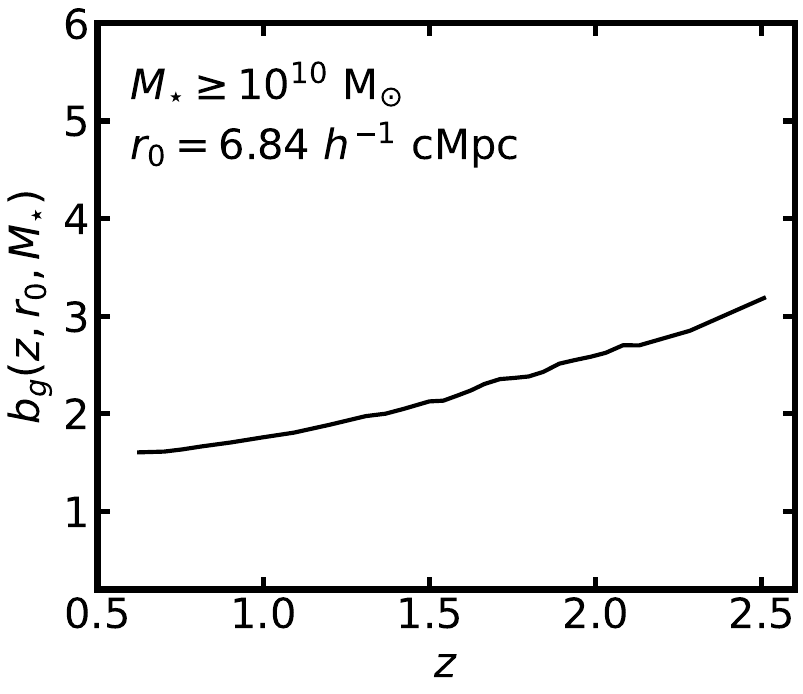}
\caption{A change of galaxy bias with redshift at fixed mass and correlation radius. Following \cite{Einasto2023}, we measure the galaxy bias at $M_\star \geq 10^{10}\ \rm{M}_\odot$, and $r_0 = 6.84\ h^{-1}\ \mathrm{cMpc}$. \label{fig:galaxybias}}
\end{figure}

We compare our results with other simulations. In Figure~\ref{fig:3dsignal}, we find that $\xi_{g+}$ has little redshift dependence. We compare our finding with the results of \cite{Tenneti2015a} and \cite{Chisari2016} who measured the projected correlation. \cite{Tenneti2015a} argued that there were not a significant redshift dependence. On the other hand, \cite{Chisari2016} reported that the evolutionary feature of the projected correlation becomes evident at small scales. However, it should be noted that the scale of interest is different from ours. We focus on $r$ from 3 to 80 $h^{-1}\ \rm{cMpc}$, which is over the scale of the one-halo alignment while their works look into shorter scale from 0.1 to 10 $h^{-1}\ \rm{cMpc}$. When looking at the correlation at large scales ($r_p>3\ h^{-1}\ \rm{cMpc}$; where $r_p$ represents the transverse separation), we find that both works do not show manifest redshift evolution of the two-point correlation. In this sense, our finding is consistent with the results of \cite{Tenneti2015a} and \cite{Chisari2016}.
The quantity shown in Figure~\ref{fig:3dsignal} is $\xi_{g+}(r) \propto b_g A_{\rm{NLA}}$, not $\xi_{m+}(r) \propto A_{\rm{NLA}}$; thus it depends on the galaxy bias, which should vary over redshifts for a fixed mass. As in Figure~\ref{fig:galaxybias}, the galaxy bias depends on redshift. The trend of redshift dependence of the galaxy bias are opposite to that of the theoretical power spectrum (equation~(\ref{eq:shapepower})), leading to a nearly constant $\xi_{g+}$ with respect to redshifts.

We also compare our findings with more recent studies. \cite{Xu2023} investigated the projected shape-density correlation of galaxies and dark-matter subhalos across different stellar mass ranges using the TNG300-1 simulation. Our results qualitatively agree with theirs, confirming that dark-matter subhalos exhibit stronger IA signal compared to galaxies. They also demonstrated that the misalignment angle depends on both stellar and halo virial masses, with the mean of misalignment angle increasing toward higher redshift. The decrease of the misalignment angles toward lower redshift agrees with our findings.

Using the TNG300-1 simulation, \cite{Rodriguez2024} investigated the anisotropic correlation function for stellar and dark-matter components of central galaxies, which is computed using pairs subtending less than 45 deg to the directions of the axes. They found that galaxies and subhalos show opposite evolutions of alignments with time. They also showed these features depend on galaxy color and halo mass. These results indicate that there are different physical processes influencing baryonic and dark-matter components, and are consistent with our findings.

\begin{figure}[!t]
\centering
\includegraphics[width=0.47\textwidth]{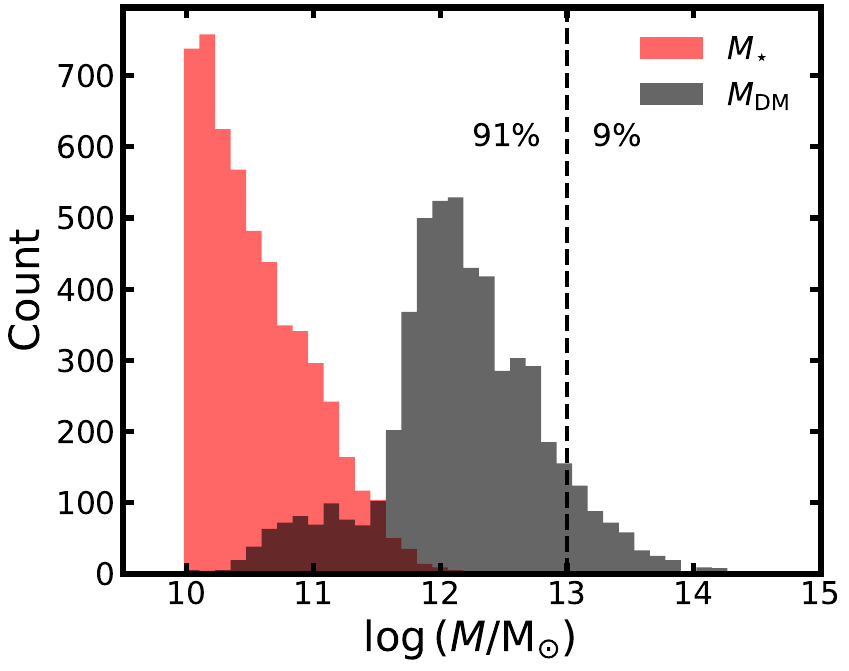}
\caption{Mass distributions of galaxy and dark-matter subhalo. The black dashed line marks a division of subhalo mass showing different trends with respect to halo mass in \cite{Kurita2021}. \label{fig:subhalomass}}
\end{figure}

We then discuss the redshift evolution of $A_{\rm{NLA}}$. \cite{Kurita2021} explored redshift dependence of dark-matter subhalos with halo mass and number density using an \textit{N}-body simulation. They showed the redshift evolution becomes more pronounced at lower redshifts (see Figure 6 therein). On the other hand, the amplitude became constant beyond $z=1.0$. The increasing feature at low redshifts is similar to our findings, while showing plateau at high redshifts for high-mass sample is in contrast with ours. In this study, we only consider the dependence of stellar mass not halo mass, which could cause the different trend. Figure~\ref{fig:subhalomass} shows the distributions of stellar mass of galaxy and dark-matter mass of subhalo. 91\% of subhalos have mass with $M_{\rm{DM}} < 10^{13}\ \rm{M}_{\odot}$, while only 9\% showing $M_{\rm{DM}} \geq 10^{13}\ \rm{M}_{\odot}$. Thus, the increase at high redshifts for dark-matter subhalos is consistent with previous results for low-mass objects. 

\cite{Tenneti2015a} presented that $A_{\rm{NLA}}$ of galaxies shows no evolution with redshift. It is similar to our results for HR5. They also showed that different samples leads to different trends, suggesting that growth of structure and dynamical processes such as galactic mergers may have played a role in this. They argued that as the misalignment is driven by such processes, the amplitude of IA should be also affected, which is partially reflected in the redshift evolution of $A_{\rm{NLA}}$.

\cite{Samuroff2021} showed $A_{\rm{NLA}}$ in Illustris, IllustrisTNG and MassiveBlack-II simulations. While the results for TNG seem no evolution, Illustris and MassiveBlack-II tend to increase with redshift. Illustris contains more late-type galaxies than IllustrisTNG \citep{Rodriguez-Gomez2019}. This is primarily due to the less effective regulation of star formation in Illustris, which contributes to this trend. They argued that because of the evolution of underlying large-scale structure and the growth of halos, the use of a fixed mass cut forced a stricter sample selection at high redshift, which excluded weakly aligned galaxies, leading to redshift evolution. This interpretation is partially consistent with ours if we consider dark-matter subhalos only. However, in our study, the IA amplitudes of galaxies do not show redshift evolution. Also, IllustrisTNG shows evident decrease of $A_{\rm{NLA}}$ at $z<0.6$, after the universe enters $\Lambda$-dominated era even considering their red galaxies of TNG only. This epoch is out of our consideration due to the limit of HR5. In this regard, our results are consistent with previous studies.

\cite{Chisari2016} investigated how $A_{\rm{NLA}}$ varies depending on the shape measurement method and the galaxy type. They did not find the redshift evolution of the IA when they measured the galaxy shapes via different inertia tensors and investigated the IA amplitude at redshift range from $z=1$ to $z=3$. At $z<1$, they also showed the similar feature to \cite{Samuroff2021}. The consistency across the considered redshift range is similar to our results, and the effect of disk galaxies on the IA amplitude shows a similar trend. 

\subsection{Redshift Evolution} \label{subsec:discuss3}

\subsubsection{Possible Origin} \label{subsubsec:discuss3-1}

Here, we discuss the possible origin of redshift dependence of the IA.
We begin with the redshift dependence of the misalignment angle. As we mentioned earlier, galaxies are more aligned with dark-matter subhalos at lower redshifts. This trend is consistent with \citet{Bhowmick2020} and \citet{Xu2023}. It is known that dynamical processes of galaxies, including the active galactic nuclei (AGN) and supernova (SN) feedback, can affect misalignment between galaxies and their dark-matter subhalos \citep{Velliscig2015a}. This misalignment is expected to influence the strength of IA of galaxies with the large-scale structure, motivating studies of accurately accounting for such discrepancy. Furthermore, decrease of the misalignment angle would permeate somehow the IA amplitude with respect to redshift.

As in Figure~\ref{fig:amp3d}, no significant redshift evolution of $A_{\rm{NLA}}$ is observed for galaxies. This result is consistent with the assumption of the NLA model, which assumes that the IA is ``frozen" at the time of galaxy formation. Dark-matter subhalos show a mild redshift evolution of $A_{\rm{NLA}}$ with a nonzero power-law slope 0.509 over $3\sigma$ from zero. During galaxy formation, the tidal field of the large-scale structure influences the distribution of collapsing matter, which leads to IAs. As the universe evolves, the large-scale structure evolves as well. When the overdense region becomes more overdense, some dark-matter subhalos undergo matter accretion that disturbs the internal structure,
and others may lose a portion of the structure because of the tidal stripping. Thus, the intrinsic shape of dark-matter subhalo changes, which might potentially weaken previously established alignments. However, galaxies are typically centered within their own dark-matter subhalos from the time of formation, a region where the gravitational potential is strongest. The gravitational potential of their dark-matter subhalos shields them somewhat from the external effects of large-scale structure evolution. In addition, the shape of stabilized early-type galaxies will be less affected by internal baryonic processes of galaxies because the star formation, which affect the shape of galaxy, can be suppressed due to the AGN, and/or SN feedback \citep[e.g.,][]{Schawinski2007,Li2018}. As discussed above, these interpretations regarding the underlying dynamical processes are consistent with the earlier suggestions made by \cite{Rodriguez2024}. Therefore, these galaxies tend to be stabilized gravitationally within their subhalos, and preserve their intrinsic shapes, which induces the decrease of the misalignment angle probability as in Figure~\ref{fig:misalignment}, \cite{Bhowmick2020}, and \cite{Xu2023}, and then breeds the lowering of the difference of $A_{\rm{NLA}}$ between galaxies and dark-matter subhalos. This fundamental difference between galaxies and dark-matter subhalos emphasizes the resultant trends in $A_{\rm{NLA}}$.

\subsubsection{Caveats} \label{subsubsec:discuss3-2}

In this work, we select early-type galaxies above a given mass. This approach includes galaxies that become massive at lower redshifts, which can introduce biases in interpreting the redshift evolution of the IA amplitude, as the sample reflects evolving galaxy population rather than tracking the evolution of individual systems. To better trace galaxy evolution, a density-limited selection may be more appropriate. We investigate such a case by fixing the sample size to match that at $z=2.5$. We find evolutionary trends in individual $A_{\rm NLA}$ values for subhalos and galaxies that are different from the case of the mass-limited sample, with slopes of $-0.193 \pm 0.170$ and $-0.462 \pm 0.376$ (see Appendix~\ref{sec:app_c} for figures).
The subhalos show relatively constant $A_{\rm NLA}$, while galaxies have higher $A_{\rm NLA}$ toward low redshifts. However, the trend in $\Delta A_{\rm{NLA}}$ and the misalignment angle distribution remains qualitatively consistent with the main results with a mass-limited sample, supporting a view that the decrease in misalignments leads to a reduction in the gap between the IA of galaxies and their host subhalos.

We calculate the strength of the IA by assuming a scenario in which the shape of a galaxy at a given point in time is determined by the density field of the matter density at that time. It is inherent in this process that comparison of the relation between the dynamical time scale of galaxies and the cosmic time scale for the tidal field. Previous observations of red galaxies have shown that this is a valid assumption \citep{Joachimi2011,Singh2015,Johnston2019,Johnston2021,Samuroff2023}. However, considering the formation and evolution of galaxies, red, elliptical, and quiescent galaxies are generally known to be the result of galaxies starting from disk galaxies and undergoing processes such as merging \citep{2010gfe..book.....M}. If the IA is determined at the time of galaxy `formation', it would imply that galaxies acquire their shape during the formation process. However, in most cases, the shape of an early-type galaxy represents the final stage of morphological evolution, where the galaxy has reached a stable state (e.g, quenched star formation, dynamical equilibrium). 

In this context, measuring the strength of the IA at each redshift is a mixed measure of the extent to which the shape of the galaxies formed at that time is related to the gravitational potential of the density field, and the extent to which galaxies formed in the past lose their IA signal through relaxation. Therefore, we should consider this complexity and the fact that the $A_{\rm NLA}$ depends on the epoch of the surrounding structure when we measure the IA amplitude.

\section{CONCLUSIONS} \label{sec:conclusion}

In this study, we investigate the IA of early-type galaxies and their dark-matter subhalos using the HR5 simulation. To do this, we classify early-type galaxies using stellar mass and kinematic information of the galaxies at each redshift. 
We measure the shape parameters using the reduced inertia tensor from the position and mass information of each stellar particle and dark-matter particle. We use two-point statistics to detect the IA correlation between galaxies (dark-matter subhalos) and the large-scale structure, and quantify the strength of the IA based on the NLA model. The main findings of our work are summarized as follows:\\

(i) We measure and compare the IA correlation function for mass-limited early-type galaxies at $z = 0.625$, 1.0, and 2.0, and find no significant redshift dependence.
The IA signal of dark-matter subhalos is consistently stronger than that of galaxies at all redshifts. 
To quantify the IA strength, we adopt the NLA model and find that galaxies show no redshift evolution in $A_{\rm IA}$, while subhalos exhibit a mild evolution, as indicated by the best-fit power-law slope.\\

(ii) By dividing the sample by stellar mass and measuring the IA strength, we confirm that IA increases with stellar mass at different redshifts. \\
Galaxies and dark-matter subhalos exhibit different IA trends with respect to the kinematic criterion based on $|V_{\rm r}/\sigma_V|$.
For galaxies, the signal weakens as the threshold increases, whereas no similar trend is observed for dark-matter subhalos.
The difference is likely because higher $|V_{\rm r}/\sigma_V|$ thresholds include a larger fraction of late-type galaxies, which weakens the IA signal,
whereas the shapes of dark-matter subhalos are largely unaffected by galaxy type.\\

(iii) The IA signal does not show significant differences depending on either the method of shape measurement (reduced or standard inertia tensor) or the direction of the projection axis. \\

(iv) We compare our findings with previous observations and simulations that show weakly evolving trends of the IA.
For galaxies, the amplitude of IA matches well with observations.
The trend of our results for galaxies is consistent with \cite{Chisari2016}, \cite{Tenneti2015a} and \cite{Samuroff2021} for the redshift range considered. 
For dark-matter subhalos, our results agree with those of \cite{Kurita2021}.\\

Our results show that the IA amplitude of galaxies does not evolve with redshift whereas that of dark-matter subhalos does. However, measured strength of the IA would depend on the modeling approach, which also incorporates how astronomical objects evolve under gravitational potential. Nevertheless, the redshift-independent feature of the IA strength of galaxies may provide useful hints about the extent to which IA may contaminate the lensing signal in future surveys.

\begin{acknowledgments}
We thank our referee for providing helpful report to enhance the manuscript.
This work is supported by the Center for Advanced Computation at Korea Institute for Advanced Study. HSH acknowledges the support of Samsung Electronic Co., Ltd. (Project Number IO220811-01945-01), the National Research Foundation of Korea (NRF) grant funded by the Korea government (MSIT), NRF-2021R1A2C1094577, and Hyunsong Educational \& Cultural Foundation. MT and SH were supported by the NRF grant (2022R1F1A1064313). J.L. is supported by the National Research Foundation (NRF) of Korea grant funded by the Korea government (MSIT, RS-2021-NR061998). C.P. and J.K. are supported by KIAS Individual Grants (PG016903, KG039603) at the Korea Institute for Advanced Study. This work benefited from the outstanding support provided by the KISTI National Supercomputing Center and its Nurion Supercomputer through the Grand Challenge Program (KSC-2018-CHA-0003, KSC-2019-CHA-0002). Large data transfer was supported by KREONET, which is managed and operated by KISTI. Y.K. is supported by KISTI under the institutional R\&D project (K25L2M2C3).
\end{acknowledgments}

\appendix
\section{SIMPLE INERTIA TENSOR} \label{sec:app_a}
\renewcommand\thefigure{\thesection.\arabic{figure}} 
\setcounter{figure}{0}
\begin{figure*}[!t]
\centering
\includegraphics[width=1\textwidth]{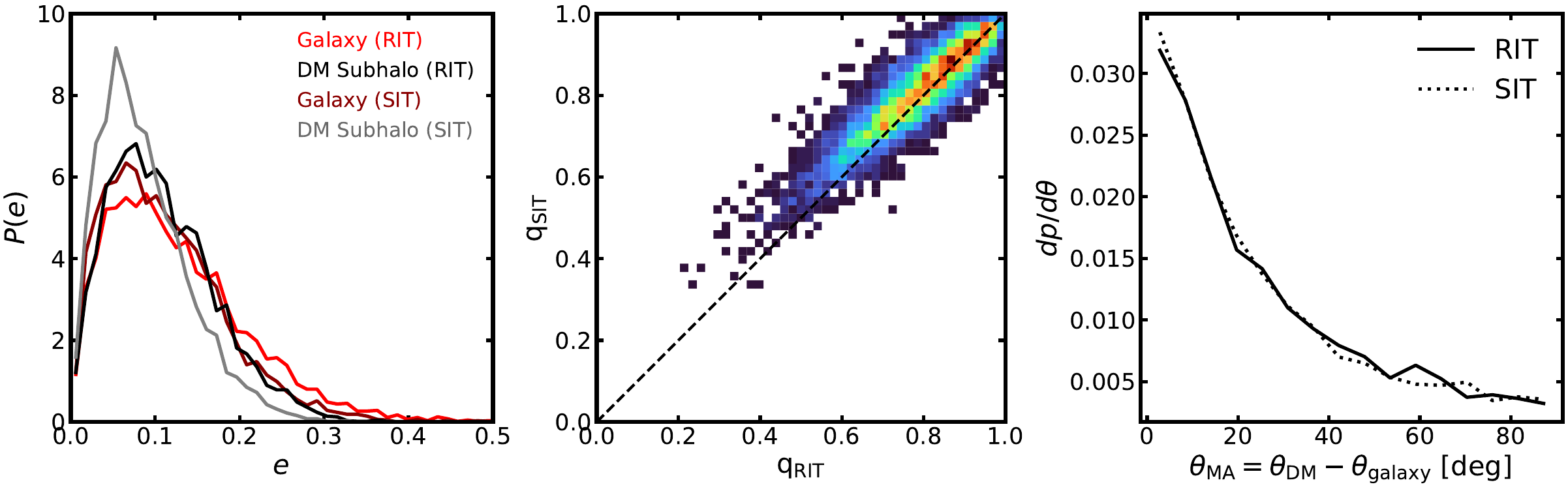}
\caption{Comparison of shape parameters obtained by simple inertia tensor to those of reduced inertia tensor. Left: total ellipticity distributions of galaxy and dark-matter subhalo. Middle: the ratio between semi-major and semi-minor axes of galaxy. Right: Misalignment angle distribution.
\label{fig:appendix_shape}}
\end{figure*}
\begin{figure*}[!t]
\centering
\includegraphics[width=1\textwidth]{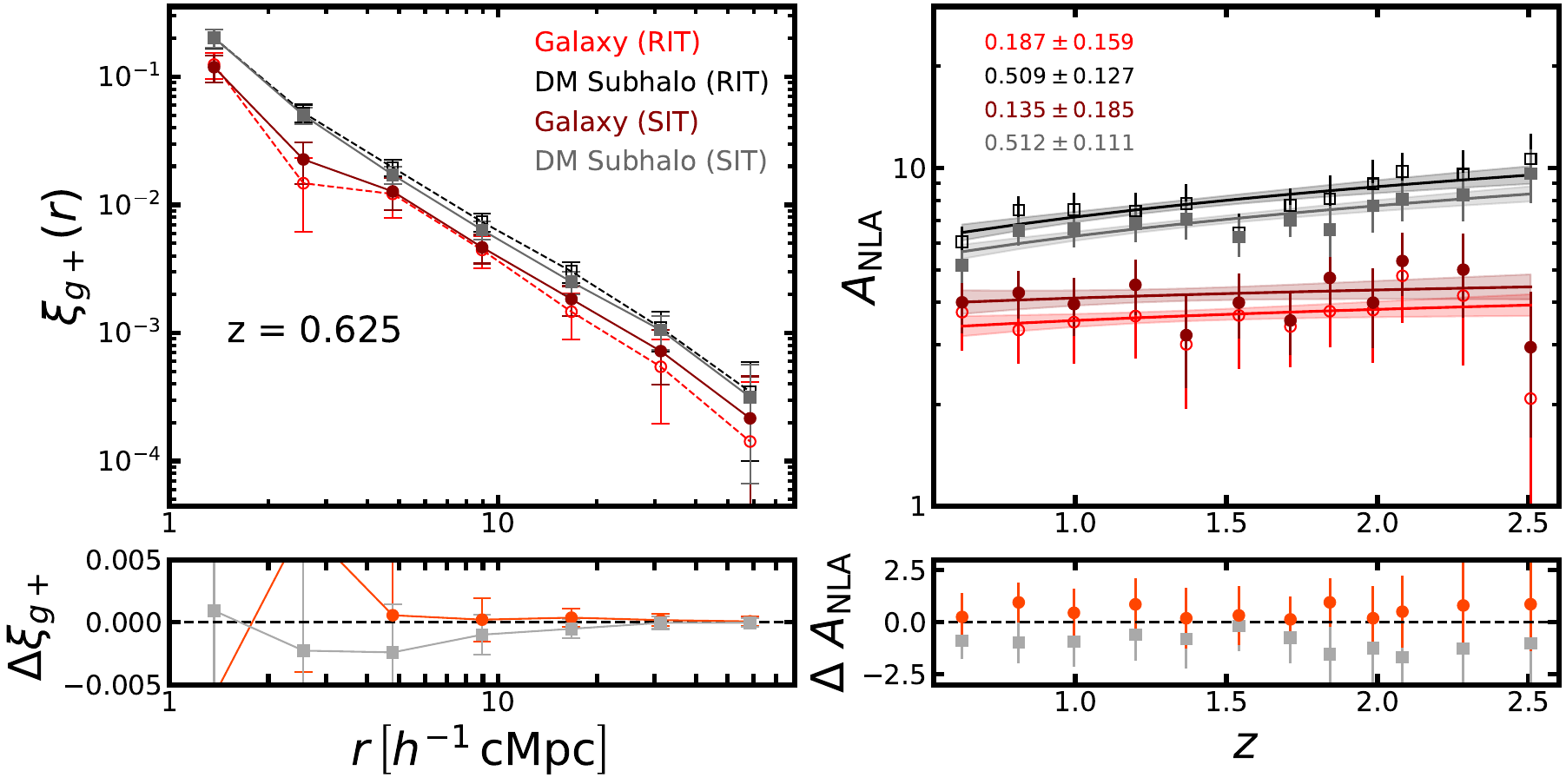}
\caption{Comparison of IA correlation and the amplitude of IA. In the top panels, the empty circles (squares) represent galaxy (dark-matter subhalo) IA correlation (top left panel) and $A_{\rm NLA}$ (top right panel) whose shapes are measured through reduced inertia tensor. The filled symbols indicate the results of simple inertia tensor. In bottom panels, orange circles (lightgrey squares) represent the difference of galaxy (dark-matter subhalo) IA correlation between simple and reduced inertia tensor.
\label{fig:appendix_IA}}
\end{figure*}

In this appendix, we investigate the effect of shape measurement method on the IA signal and amplitude. We first compare the shape parameters obtained by simple (unweighted) inertia tensor to those of reduced inertia tensor. To do this, we define the simple inertia tensor as follows:
\begin{equation} \label{eq:SIT}
I_{ij} = \frac{1}{M} \sum^n_{k=1}{m_k x_{k,i}x_{k,j}}
\end{equation}
The terms in equation~\ref{eq:SIT} are identical to equation~\ref{eq:inertia}. We evaluate the eigenvectors and eigenvalues of the tensor, and estimate the shape parameters.

Figure~\ref{fig:appendix_shape} shows comparison of shape parameters obtained by reduced and simple inertia tensor. In the left and middle panels, we find that the shapes with reduced inertia tensor are more elliptical than those obtained with simple inertia tensor. This is in contrast with the results of \cite{Chisari2015}. However, we note that our definition of the ellipticity is different with the definition that they used. The definition of the weights, $r^2_n$, is also different from what we use. They define $r^2_n$ the three-dimensional distance of the stellar particle to the center of galaxy as the spherically symmetric weights, while we utilize the elliptical weights. We also note that \cite{Chisari2015} did not classify the galaxy when they analyzed the ellipticity of galaxy. The right panel displays the misalignment angle distribution, which are similar to each other.

Figure~\ref{fig:appendix_IA} displays IA correlation and the amplitude of IA. In the left panels, the IA correlation does not show a significant dependence on the shape measurement. 
In the right panels, $A_{\rm{NLA}}$ shows similar trend to the left. Galaxy (dark-matter subhalo) whose shape is obtained by simple inertia tensor displays very slightly higher (lower) amplitude than reduced inertia tensor case. However, considering the difference, it is not statistically significant and does not change the overall trend. Our test shows inconsistent results against \cite{Chisari2016}, which compare the effect of shape measurement method (Figure 11 therein). Their results show that the amplitude of IA for elliptical galaxy whose shape is obtained by simple inertia tensor is much larger than reduced inertia tensor case. However, as we noted, the scale of the correlation measurement is different from ours, and the weights imposed in the reduced case are different. Also, the difference shown in bottom panels is similar to zero, so that the systematics of shape measurement does not affect the trend of redshift evolution.

\section{GRID-LOCKING} \label{sec:app_b}
\renewcommand{\thefigure}{\thesection.\arabic{figure}}
\setcounter{figure}{0}

In this appendix, we test the possibility of the contamination of the `grid-locking' effect in our results. The simulations based on AMR technique such as RAMSES tend to show that the spins of galaxy are aligned with the simulation grid vector \citep{Hahn2010,Dubois2014,Codis2015a}. This is a well-known issue that the grid-locking effect occurs in cartesian base Poisson solver when calculating the force from gravitational potential \citep{hockney2021computer,May1984}. In particular, the tendency for the spin to be artificially aligned with the grid vector seems better seen in low-mass galaxies, and in galaxies with lower redshifts \citep{Hahn2010,Danovich2012}. Because the spin of a galaxy is one tracer of its orientation, the fact that the spin shows numerical errors due to the grid locking effect suggests the possibility that it affects the orientation of the galaxy and, by extension, biases the IA of the galaxy. 
\begin{figure}[!t]
\centering
\includegraphics[width=0.47\textwidth]{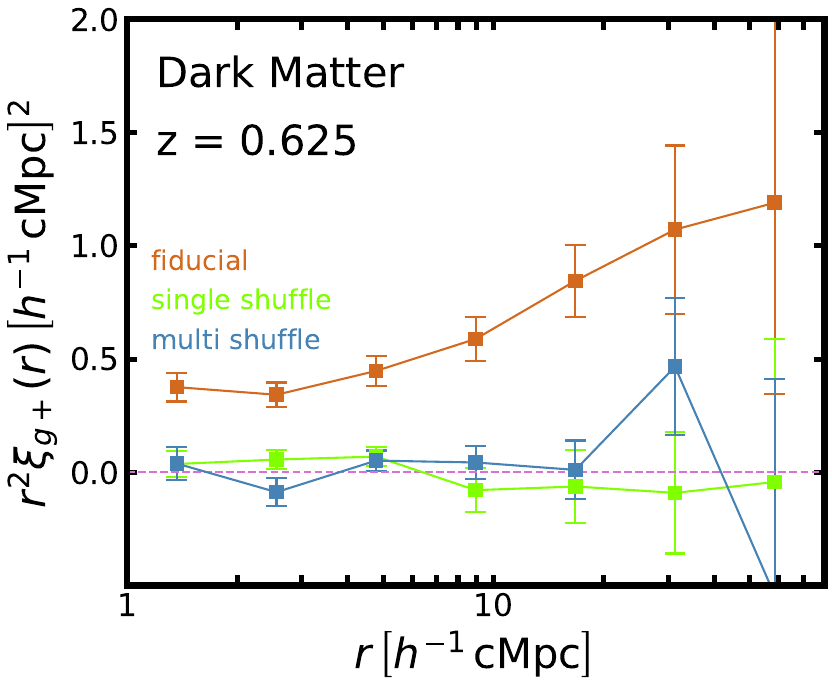}
\caption{The correlation function to examine grid-locking effect. The ellipticity and position angle are randomly mixed, for once and for several times. The position of dark-matter subhalos remains without shuffle. We multiply $\xi(r)$ by $r^2$ for presentation purpose. We only show the correlation of dark-matter subhalo for clarity, but we confirm the correlation of galaxies is consistent with dark-matter subhalos. Shuffled results indicate that shape of galaxy and dark-matter subhalo are not affected by the grid-locking effect.}
\label{fig:appendix_grid-locking}
\end{figure}

We therefore verify whether the IA signal of HR5 galaxies is likely to be contaminated by these phenomena. To do this, we adopt the method of \cite{Codis2015a}. Because we are interested in the alignment between the shape and the density field, not the spin, we compute the same statistics by randomly permuting the shape parameters ellipticity and position angle while keeping the density field fixed by keeping galaxy position unchanged. The random permutation removes the physical two-point correlation signal. If the IA signal is biased by grid-locking, then random shuffling of the shape will not change the number or orientation of the grid-locked objects, so the correlation will appear regardless of the permutation. Figure~\ref{fig:appendix_grid-locking} shows the results of measuring the IA correlation for single-shuffled, and multi-shuffled cases with the original IA signal at $z=0.625$. The two-point correlation is consistent with the null (zero) no matter how many times we shuffle. We confirm that the shape parameters of galaxy were not affected by the grid locking effect and that the IA signal we measured has a physical origin.

\section{$A_{\rm NLA}$ FOR DENSITY-LIMITED SAMPLES} \label{sec:app_c}
\renewcommand{\thefigure}{\thesection.\arabic{figure}}
\setcounter{figure}{0}

\begin{figure}[!ht]
\centering
\includegraphics[width=0.47\textwidth]{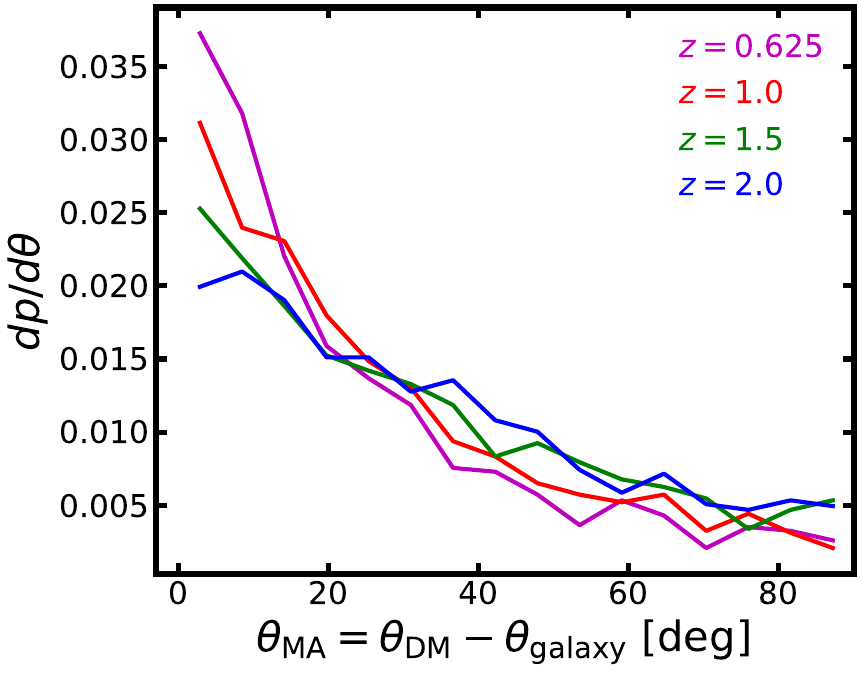}
\caption{Similarly to Figure~\ref{fig:misalignment}, the distribution of misalignment angles for density-limited samples evolves with redshift.}
\label{fig:appendix_maa_denlim}
\end{figure}

\begin{figure}[!ht]
\centering
\includegraphics[width=0.47\textwidth]{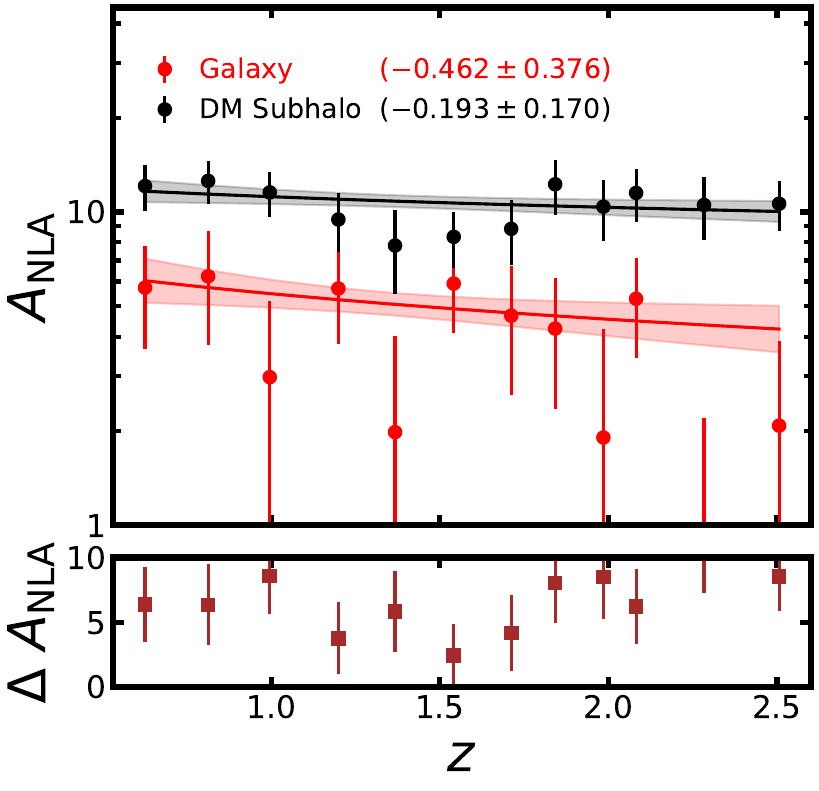}
\caption{Similar to Figure~\ref{fig:amp3d}, but for density-limited samples.}
\label{fig:appendix_amp_denlim}
\end{figure}

In this appendix, we display the redshift dependence of $A_{\rm NLA}$ for the density-limited samples discussed in Section~\ref{subsubsec:discuss3-2}. We select $\sim 1300$ massive early-type galaxies at $z=2.5$ and keep this sample density at low redshifs. Figure~\ref{fig:appendix_maa_denlim} shows the distribution of the misalignment angle for the density-limited case. Figure~\ref{fig:appendix_amp_denlim} exhibits the redshift evolution of $A_{\rm{NLA}}$ for this case. 

\bibliography{Delphi_v8}{}
\bibliographystyle{aasjournalv7}



\end{document}